\documentclass[twoside,twocolumn,9pt]{article}
\usepackage{extsizes}
\usepackage[super,sort&compress,comma]{natbib} 
\usepackage[version=3]{mhchem}
\usepackage[left=1.5cm, right=1.5cm, top=1.785cm, bottom=2.0cm]{geometry}
\usepackage{balance}
\usepackage{times,mathptmx}
\usepackage{sectsty}
\usepackage{graphicx} 
\usepackage{lastpage}
\usepackage[format=plain,justification=justified,singlelinecheck=false,font={stretch=1.125,small,sf},labelfont=bf,labelsep=space]{caption}
\usepackage{float}
\usepackage{fancyhdr}
\usepackage{fnpos}
\usepackage[english]{babel}
\addto{\captionsenglish}{%
  
}
\usepackage{array}
\usepackage{droidsans}
\usepackage{charter}
\usepackage[T1]{fontenc}
\usepackage[usenames,dvipsnames]{xcolor}
\usepackage{setspace}
\usepackage[compact]{titlesec}
\usepackage{hyperref}
\usepackage{bigints}
\usepackage{braket}

\usepackage{epstopdf}
\usepackage[normalem]{ulem}

\definecolor{cream}{RGB}{222,217,201}

\newcommand{\eqn}[1]{Eqn.~\eqref{#1}} 
\newcommand{\onlinecite}{\citenum}

\newcommand{\eu}{\mathrm{e}^}
\newcommand{\rmd}{\mathrm{d}}

\begin{document}

\pagestyle{fancy}
\fancypagestyle{plain}{

\fancyhead[C]{\includegraphics[width=18.5cm]{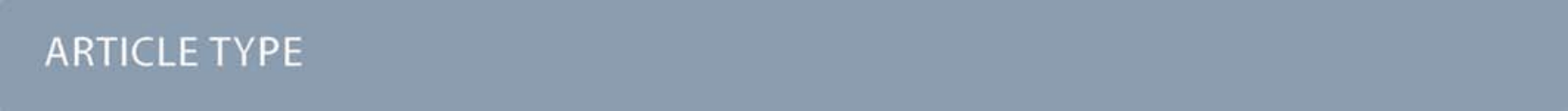}}
\fancyhead[L]{\hspace{0cm}\vspace{1.5cm}\includegraphics[height=30pt]{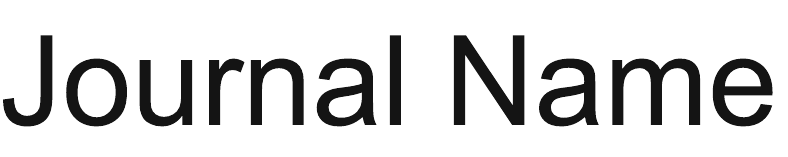}}
\fancyhead[R]{\hspace{0cm}\vspace{1.7cm}\includegraphics[height=55pt]{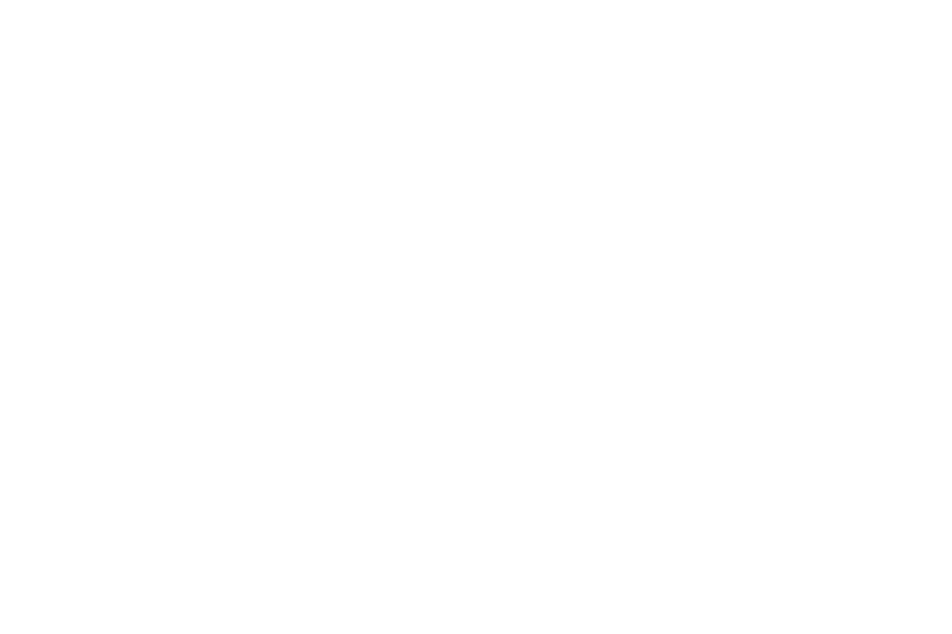}}
\renewcommand{\headrulewidth}{0pt}
}

\makeFNbottom
\makeatletter
\renewcommand\LARGE{\@setfontsize\LARGE{15pt}{17}}
\renewcommand\Large{\@setfontsize\Large{12pt}{14}}
\renewcommand\large{\@setfontsize\large{10pt}{12}}
\renewcommand\footnotesize{\@setfontsize\footnotesize{7pt}{10}}
\makeatother

\renewcommand{\thefootnote}{\fnsymbol{footnote}}
\renewcommand\footnoterule{\vspace*{1pt}%
\color{cream}\hrule width 3.5in height 0.4pt \color{black}\vspace*{5pt}} 
\setcounter{secnumdepth}{5}

\makeatletter 
\renewcommand\@biblabel[1]{#1}            
\renewcommand\@makefntext[1]%
{\noindent\makebox[0pt][r]{\@thefnmark\,}#1}
\makeatother 
\renewcommand{\figurename}{\small{Fig.}~}
\sectionfont{\sffamily\Large}
\subsectionfont{\normalsize}
\subsubsectionfont{\bf}
\setstretch{1.125} 
\setlength{\skip\footins}{0.8cm}
\setlength{\footnotesep}{0.25cm}
\setlength{\jot}{10pt}
\titlespacing*{\section}{0pt}{4pt}{4pt}
\titlespacing*{\subsection}{0pt}{15pt}{1pt}

\fancyfoot{}
\fancyfoot[LO,RE]{\vspace{-7.1pt}\includegraphics[height=9pt]{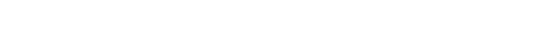}}
\fancyfoot[CO]{\vspace{-7.1pt}\hspace{13.2cm}\includegraphics{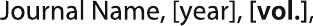}}
\fancyfoot[CE]{\vspace{-7.2pt}\hspace{-14.2cm}\includegraphics{RF.pdf}}
\fancyfoot[RO]{\footnotesize{\sffamily{1--\pageref{LastPage} ~\textbar  \hspace{2pt}\thepage}}}
\fancyfoot[LE]{\footnotesize{\sffamily{\thepage~\textbar\hspace{3.45cm} 1--\pageref{LastPage}}}}
\fancyhead{}
\renewcommand{\headrulewidth}{0pt} 
\renewcommand{\footrulewidth}{0pt}
\setlength{\arrayrulewidth}{1pt}
\setlength{\columnsep}{6.5mm}
\setlength\bibsep{1pt}

\makeatletter 
\newlength{\figrulesep} 
\setlength{\figrulesep}{0.5\textfloatsep} 

\newcommand{\topfigrule}{\vspace*{-1pt}%
\noindent{\color{cream}\rule[-\figrulesep]{\columnwidth}{1.5pt}} }

\newcommand{\botfigrule}{\vspace*{-2pt}%
\noindent{\color{cream}\rule[\figrulesep]{\columnwidth}{1.5pt}} }

\newcommand{\dblfigrule}{\vspace*{-1pt}%
\noindent{\color{cream}\rule[-\figrulesep]{\textwidth}{1.5pt}} }

\makeatother

\twocolumn[
  \begin{@twocolumnfalse}
\vspace{3cm}
\sffamily
\begin{tabular}{m{4.5cm} p{13.5cm} }

\includegraphics{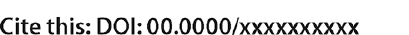} & \noindent\LARGE{\textbf{Revisiting nuclear tunnelling in the aqueous ferrous--ferric electron transfer}}\\
\vspace{0.3cm} & \vspace{0.3cm} \\

 & \noindent\large{Wei Fang,$^{a,\dag}$ Rhiannon A. Zarotiadis,$^{a,\dag}$ and Jeremy O. Richardson$^{a,\ast}$} \\

\includegraphics{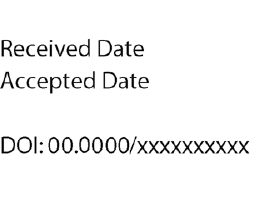} & \noindent\normalsize{%
The aqueous ferrous--ferric system provides a classic example of an electron-transfer process in solution.
There has been a long standing argument spanning more than three decades around the importance of nuclear tunnelling in this system,
with estimates based on  Wolynes theory suggesting a quantum correction factor of 65, while estimates based on a related spin-boson model suggest a smaller factor of 7--36.
Recently, we have shown that Wolynes theory can break down for systems with multiple transition states leading to an overestimation of the rate, and we suggest that a liquid system such as the one investigated here may be particularly prone to this. 
We re-investigate this old yet interesting system with the first
application of the recently developed golden-rule quantum transition-state theory (GR-QTST). 
We find that GR-QTST can be applied to this complex system without apparent difficulties
and that it gives a prediction for the quantum rate 6 times smaller than that
from Wolynes theory.
The fact that these theories give different results suggests that
although it is well known that the system can be treated using linear response and therefore resembles a spin-boson model in the classical limit, this approximation is questionable in the quantum case. 
It also intriguingly suggests the possibility that
the previous predictions were overestimating the rate due to a break down of Wolynes theory.
}

\end{tabular}

 \end{@twocolumnfalse} \vspace{0.6cm}

  ]

\renewcommand*\rmdefault{bch}\normalfont\upshape
\rmfamily
\section*{}
\vspace{-1cm}

\footnotetext{\textit{$^{a}$~Laboratory of Physical Chemistry, ETH Z\"urich, 8093 Z\"urich, Switzerland}}
\footnotetext{\textit{$^{\dag}$~These authors contributed equally}}
\footnotetext{\textit{$^{\ast}$~E-mail: jeremy.richardson@phys.chem.ethz.ch}}
 
\section{Introduction} \label{intro}

The realms of chemistry and biology serve us with a colourful variety of reactions affected by nuclear tunnelling. \cite{BellBook}
In chemistry, tunnelling is predicted to be important under a wide range of conditions
from astrochemical reactions occurring on cosmic dust \cite{astrochemistry2} 
and nuclear fusion in stars \cite{astrochemistry1} 
to organic chemistry, where even heavy-atom tunnelling
has been identified. \cite{orgchemistry, orgchemistry2}
Biological systems have also been suspected of employing nuclear tunnelling, for instance in photosynthesis taking place in bacteria \cite{tunneling1} or during 
enzyme catalysis. \cite{EnzymeCatalysisNucTun1, EnzymeCatalysisNucTun2}
To resolve such controversial hypotheses, a reliable method to calculate effects of nuclear tunnelling is clearly desirable. Such a theory will be useful to quantify the relevance of tunnelling in a given reaction.

In this work we focus specifically on the case of electron-transfer reactions. \cite{Marcus1964review, Marcus1985review, Marcus1993review}
These reactions are nonadiabatic and governed by a change of electronic state and one cannot therefore employ the Born--Oppenheimer approximation. \cite{ChandlerET}
The rate is however well described by Fermi's golden rule,\cite{FermiBook,Nitzan}
although in practice this cannot be evaluated for complex molecular systems as it requires complete knowledge of the internal eigenstates of the system.
The simplest approach is to map the system onto a harmonic spin-boson model, for which the rate can be evaluated exactly.
The mapping is of course not exact,
and thus this procedure involves an uncontrolled approximation.

Modern quantum rate theories \cite{Lawrence2020rates} are typically based on 
the path-integral approach to quantum mechanics, \cite{Feynman} which allows tunnelling and other nuclear quantum effects (NQEs) to be included efficiently into molecular simulations \cite{Markland2018review} using a quantum-classical correspondence. \cite{Chandler+Wolynes1981} However, because the rate is not defined as a simple expectation value of the density matrix, but rather in terms of a time correlation function, \cite{Miller1998rate} it is by no means trivial to calculate rates in this way and further approximations are required.
In this paper, we will concentrate in particular on quantum transition-state theories and not consider dynamical methods. \cite{Kananenka2018rate,Makri2015QCPI}

Semiclassical instanton rate theory \cite{Miller1975semiclassical, Uses_of_Instantons_PCCP, Perspective} predicts the tunnelling rate and mechanism via locating the optimal tunnelling pathway (the instanton) defined by a stationary-action principle.
Based on a similar first-principles derivation as in the normal regime, \cite{AdiabaticGreens,InstReview} instanton theory has been extended to treat electron-transfer reactions \cite{GoldenGreens,GoldenRPI,AsymSysBath}
in both the normal and inverted regimes.\cite{inverted}
It has the most rigorous derivation of the methods discussed in this paper,
shows excellent agreement with exact methods on model systems and is well suited for gas-phase electron-transfer reactions.
However, for liquid systems, it is formally not valid to apply instanton theory, \cite{InstReview} although in some cases approximate application is possible by using an implicit solvent model or by freezing all atoms not expected to be involved in tunnelling at the transition state (TS) geometry. \cite{Rommel2012enzyme}
For the general case, an extension of instanton theory that allows for sampling is desired.

Wolynes theory \cite{wolynes} is an approximate quantum rate theory
which describes electron transfer in Fermi's golden-rule limit.
It is defined in terms of path integrals
which can be evaluated using an $N$-bead discretization with each bead assigned to either the reactant or product electronic state.
The method of path-integral molecular dynamics (PIMD) \cite{Parrinello1984Fcenter} opens Wolynes theory up to the sampling tools of molecular dynamics (MD) calculations and accordingly makes it a computationally feasible approach for simulating atomistic systems.\cite{Bader, wolynes-app1, wolynes-app2}
Lawrence and Manolopoulos have recently shown that Wolynes theory can also be successfully extrapolated to the Marcus inverted regime. \cite{Lawrence2018Wolynes} 

Wolynes theory has been thoroughly investigated not only for atomistic but also for model systems
such as the spin-boson model, where it compares very well to exact results,\cite{Bader,Cao1997nonadiabatic} because it recovers the stationary phase-approximation. \cite{Weiss}
A limitation to Wolynes theory however is, that it does not tend to the classical limit for anharmonic systems. \cite{GoldenRPI,GRQTST}
Recently, we have also pointed out another crucial limitation of Wolynes theory, which is that its approximations break down when a system consists of two or more transition states.\cite{GRQTST2} This break-down can manifest itself as an overestimation of the reaction rate by more than an order of magnitude in either the classical or the quantum limit.
This may lead to the prediction of an artificial tunnelling factor.
The break-down of Wolynes theory can be related to its lack of connection to instanton theory, as it is observed that it does not necessarily sample paths close to the diabatic crossing seam, where the instantons are located, but can rather include unphysical configurations far from the seam. 
This makes any mechanistic insight or a correct rate prediction impossible.

The quantum-instanton method \cite{Miller2003QI} suffers in a similar way when applied to strongly asymmetric barriers, which can be explained from an analysis in terms of semiclassical pathways and corrected by introducing a projection to connect it to the instanton.\cite{QInst}
A further example to back this line of argumentation is the success of ring-polymer molecular dynamics (RPMD), \cite{RPMD1, RPMD2, RPMD3} which was shown to be closely connected to the semiclassical instanton rate theory in the deep-tunnelling regime. \cite{RPInst}
Standard RPMD rate theory is only applicable in the adiabatic limit, \cite{succRPMD1, succRPMD2, RosanaOnAzzouz-Borgis} but has also been used to study electron tunnelling (instead of NQEs) in the aqueous ferrous--ferric system. \cite{Menzeleev2011ET}
Building on the success of adiabatic RPMD rate theory, attempts were made to extend it to treat the nonadiabatic limit.
Two such attempts are the kinetically-constrained RPMD \cite{Menzeleev2014kinetic, Kretchmer2016KCRPMD, Kretchmer2018KCRPMD}
and the isomorphic RPMD method, \cite{Tao2018isomorphic,Tao2019RPSH} which 
do not always give reliable tunnelling factors.\cite{Lawrence2019ET, Lawrence2019isoRPMD}
One can in turn relate this behaviour to their lack of connection to instanton theory.

We therefore proposed golden-rule quantum transition-state theory (GR-QTST) \cite{GRQTST} in order to overcome issues of possible break-down behaviour by keeping a relation to instanton theory, but also retain the advantageous feature of Wolynes theory which includes not only the instanton but also paths in its vicinity.
This method is computed in a similar way to Wolynes theory, except that a constraint is imposed on the sampled paths such that the energy on the reactant and product states must match.
Adding such a constraint has been proposed as a general approach for defining quantum transition-state theories. \cite{nonoscillatory}
This constraint is automatically obeyed by all instantons, which ensures a strong connection to instanton theory, and it also retains the correct classical limit.
GR-QTST has been shown to perform very well for model systems in both the normal and inverted regimes,\cite{GRQTST}
including the multidimensional spin-boson model.
GR-QTST was also investigated for systems with multiple transition states, where Wolynes theory breaks, and provides accurate rate predictions. \cite{GRQTST2}
For the systems tested so far, we claimed it was the most accurate imaginary-time path-integral method currently available.
However, GR-QTST has not previously been applied to atomistic simulations.
This work therefore aims to investigate the applicability of GR-QTST as well as to see what physical insights it can offer by revisiting the early papers on the aqueous ferrous--ferric electron transfer. \cite{Kuharski, Bader, Marchi1991tunnelling}
This is a prototypical atomistic system for which a computationally inexpensive force field is readily available.\cite{Kuharski}
The seemingly simple interactions in this system forge a rough, high-dimensional, anharmonic potential energy surface (PES), and display high levels of complexity due to it being atomistic, which is more realistic and complex far beyond any of the models previously studied by GR-QTST.

\begin{figure}
    \centering
    \includegraphics[width=0.25\textwidth]{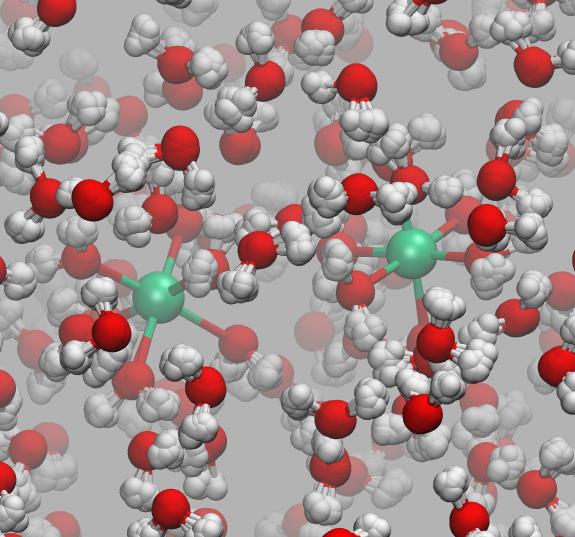}
    \caption{Snapshot of the aqueous ferrous--ferric system from a PIMD trajectory.
    The \ce{Fe^2+} and \ce{Fe^3+} ions are shown in green and are solvated in an octahedral ligand environment.
    }
    \label{snapshot}
\end{figure}

The system is depicted in Fig.~\ref{snapshot}, and despite its seemingly innocent appearance, there has been a long standing argument over the importance of nuclear tunnelling in this system at room temperature.
The quantum correction factor reported in Ref. \citenum{Bader} and calculated based on similar ideas to Wolynes theory is approximately 65, suggesting a significant contribution from nuclear tunnelling. 
This estimate is significantly larger ($\sim$~6 times) than other predictions made at the time. \cite{Song1993quantum2}
Due to this discrepancy in the calculated tunnelling enhancements, the aqueous ferrous--ferric system is a good atomistic test case worth revisiting with newly developed rate theories.
It is also of interest to investigate the applicability of Wolynes theory in this case
in view of a possible overestimation of the nuclear tunnelling effect.

In order to reexamine the earlier findings of Chandler and co-workers \cite{Kuharski, Bader} and add the investigation of GR-QTST for this system, we recapitulate the various rate theories under study in Section \ref{method}. 
The computational details of the implementation of each rate theory are given in Section \ref{compdet} and the results are presented and discussed in Section \ref{results}.
We conclude in Section \ref{conclusion} on the quality and appropriateness of the various quantum rate theories and discuss our work in the context of that of others.

\section{Theory} \label{method}

The quantum Hamiltonian describing an electron-transfer reaction is \cite{ChandlerET}
\begin{align}
\hat{H}=\hat{H}_0|0\rangle\langle0| + \hat{H}_1|1\rangle\langle1| + \Delta(|0\rangle\langle1|+|1\rangle\langle0|),
\end{align}
in which $\Delta$ is the electronic coupling,
and $\hat{H}_n = \sum_{j=1}^D\hat{p}_j^2/2m_j + V_n(\hat{\mathsf{x}})$ is the nuclear Hamiltonian for the electronic state $\ket{n}$
with the PES $V_n(\mathsf{x})$,
where $\mathsf{x}=(x_1,...,x_D)$ is the nuclear geometry and the index $j$ runs over each 
of the $D$ nuclear degrees of freedom of the system with momentum $p_j$ and associated mass $m_{j}$.
Following the work of Chandler and co-workers, \cite{Bader}
$\Delta$ is assumed to be a constant, which is known as the Condon approximation.

The rate, in the limit of small $\Delta$, is in principle given by Fermi's golden rule, \cite{FermiBook}
which is commonly approximated using Marcus theory.
\cite{Marcus1985review}
For a symmetric system, the rate is then given by a simple equation:
\begin{equation}
k_{\text{Marcus}} = \frac{\Delta^2}{\hbar}  \sqrt{\frac{\pi\beta}{\Lambda}}\eu{-\beta\Lambda/4},  
\label{Marcus_R}
\end{equation}
where $\Lambda$ is the reorganisation energy, defined by the average energy gap between the two PESs for a classical ensemble in the reactant state.
Despite its successful use in a wide range of applications, it ignores NQEs.
Various methods for evaluating electron-transfer rates including these quantum effects in the golden-rule limit have been derived,
typically based on a spin-boson model of the system Hamiltonian.
\cite{Siders1981quantum,Bixon1999review_PCCP,Nitzan}
However in this paper, we shall focus on methods based on the imaginary-time path-integral formulation,
which are applicable for complex molecular systems described by atomistic Hamiltonians.

\begin{figure}
    \centering
    \includegraphics[width=0.6\columnwidth]{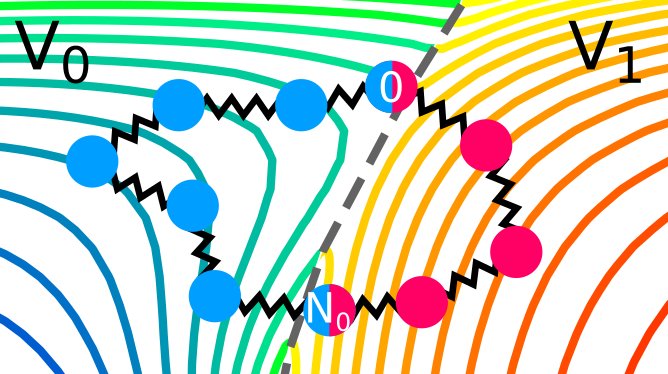}
    \caption{
    Illustration of a ring polymer on two diabatic PESs $V_0$ and $V_1$ in a two-dimensional nuclear configurational space.
    Only contours for the lowest PES are shown at any configuration.
    The blue (red) beads of the ring polymer represent the imaginary-time path on the reactant (product) electronic state.
    In this example, $N_0=6$ and $N_1=4$ giving a total of $N=10$ beads.
    }
    \label{rp2}
\end{figure}

In electron-transfer theory,
cyclic paths are formed by joining together an open-ended path
on the reactant state and an open-ended path on the product state.
For a ring-polymer representation of these paths, we introduce $\lambda$ as a dimensionless order parameter, 
which determines how the ring-polymer beads are distributed between the two diabatic states $\ket{0}$ and $\ket{1}$. It is defined by
\begin{equation} \label{lambda}
1 - \lambda
\equiv 1 - \frac{N_1}{N} \equiv \frac{N_0}{N},
\end{equation}
where $N_0$, $N_1$ and $N$ are integers according to the discrete distribution of beads, i.e.\ $N_0$ on the diabatic state $\ket{0}$ and $N_1$ on the diabatic state $\ket{1}$. The total number of ring-polymer beads is $N$.
An illustration of such a ring polymer is given in Fig.~\ref{rp2}.
The two extreme distributions assign all beads to just one diabatic state can be described with the order parameter $\lambda = 0$ for the case that all beads are on the reactant PES, $V_0$, and $\lambda = 1$ for the case that all beads are on the product PES, $V_1$.
The unconstrained ensemble of ring polymers can be sampled using thermostatted PIMD based on the following extended Hamiltonian:
\begin{subequations}
	\label{HRP2}
\begin{align}
    H_{\text{RP}}^{(\lambda)} &= \sum_{i=1}^{N}\sum_{j=1}^{D}\frac{[p_j^{(i)}]^2}{2m_j} +U_{\text{RP}}(\mathbf{x}) + U_N^{(\lambda)}(\mathbf{x}) \label{HRP}\\
    U_\text{RP}(\mathbf{x}) &= \sum_{i=1}^{N}\sum_{j=1}^{D}\frac{1}{2}m_j\omega_N^2[x_j^{(i)}-x_j^{(i-1)}]^2 \\
    U_N^{(\lambda)}(\mathbf{x}) &= \sum_{i=1}^{N_0-1}V_0(\mathsf{x}^{(i)})+\sum_{i=N_0+1}^{N-1}V_1(\mathsf{x}^{(i)}) \nonumber \\
    &+\sum_{i\in\{N_0,N\}}\frac{1}{2}[V_0(\mathsf{x}^{(i)})+V_1(\mathsf{x}^{(i)})], \quad \text{for $0<\lambda<1$} 
	 \\
    U_N^{(0)}(\mathbf{x}) &= \sum_{i=1}^{N}V_0(\mathsf{x}^{(i)}),
	 \\
    U_N^{(1)}(\mathbf{x}) &= \sum_{i=1}^{N}V_1(\mathsf{x}^{(i)}),
\end{align}
\end{subequations}
where $\omega_N=1/\beta_N\hbar$ with $\beta_N=\beta/N$ and $\beta=1/k_\mathrm{B}T$, $k_\mathrm{B}$ is the Boltzmann constant, $T$ is the temperature 
and $\textbf{x}=\{\mathsf{x}^{(1)},\dots,\mathsf{x}^{(N)}\}$ 
are the positions of the beads with conjugate momenta $\mathbf{p}$.
The cyclic index $i$ runs over each bead such that $\mathsf{x}^{(0)}\equiv \mathsf{x}^{(N)}$.
Unconstrained free energies are defined in terms of the ensemble of ring polymers by
\begin{align}
    \eu{-\beta F_\text{u}(\lambda)} = \frac{1}{(2\pi\hbar)^{ND}} \iint \eu{-\beta_N H_{\text{RP}}^{(\lambda)}} \, \text{d}\textbf{x} \, \text{d}\textbf{p}
\end{align}
and the reactant free energy by $F_0 = F_\text{u}(0)$.

First we define Wolynes rate theory, \cite{Wolynes1987nonadiabatic}
which was derived via a second-order cumulant approximation of the time-correlation function \cite{Miller1983rate} and is given by
\begin{equation}
\label{Wolynes_R}
k_{\text{Wolynes}} = \frac{\Delta^2}{\hbar}\sqrt{2\pi\beta}\left(-\frac{\rmd^2 F_\text{u}}{\rmd\lambda^2}\right)_{\lambda=\lambda^{\ast}}^{-\frac{1}{2}} \eu{-\beta (F_{\text{u}}(\lambda^{\ast})-F_0)},
\end{equation}
where $F_{\text{u}}(\lambda^{\ast})$ is the maximum unconstrained free energy with respect to the order parameter $\lambda$. For symmetric systems, the maximum occurs at $\lambda^{\ast}=0.5$.

The GR-QTST method approximates the golden-rule rate using the ansatz:\cite{GRQTST}
\begin{equation}
	\label{grqtst}
	k_\text{GR-QTST} = \frac{2\pi\beta\Delta^2}{\hbar} \, \eu{-\beta (F_\text{c}(\lambda^{\ast})-F_0)},
\end{equation}
in which $\lambda^{\ast}$ is the same as that used in Wolynes theory, i.e.\ the maximum of the unconstrained free energy according to the approach introduced by us in Ref.~\onlinecite{GRQTST2}.
$F_\text{c}(\lambda)$ is the free energy under the constraint $\sigma_{\lambda}\!(\textbf{x})=0$ given by
\begin{equation}
    \eu{-\beta F_{\text{c}}(\lambda)} = \frac{1}{(2 \pi \hbar)^{ND}} \iint \eu{- \beta_{N} H_{\text{RP}}^{(\lambda)}} \delta(\sigma_\lambda) \, \rmd\textbf{x} \, \rmd \mathbf{p},
\end{equation}
with the ring-polymer Hamiltonian $H_{\text{RP}}^{(\lambda)}$ defined by Eqn.~\eqref{HRP} and the constraint function defined according to the ansatz of Ref.~\citenum{GRQTST} as $\sigma_\lambda\!(\mathbf{x})=\frac{2}{3}\beta(E_0^\text{v}-E_1^\text{v})$.
The virial energy estimators of the product and reactant paths, $E_0^\text{v}$ and $E_1^\text{v}$, are defined as in Ref.~\citenum{GRQTST2}
and are functions of the potentials and gradients of the beads corresponding to one particular state.
The constraint is designed to enforce energy conservation
for the ring polymers sampled in the simulation,
which is known to give a strong connection to quantum transition-state theories. \cite{nonoscillatory}

The classical rate in the golden-rule limit is defined by \cite{nonoscillatory, ChandlerET}
\begin{equation}
k_{\text{cl}}
    = \frac{2\pi\beta\Delta^2}{\hbar} \, \eu{-\beta (F_\text{c}^{\text{cl}}-F_0^{\text{cl}})},
\label{kcl}
\end{equation}
where $F_0^{\text{cl}}$ is the classical free energy of the reactant, and $F_\text{c}^{\text{cl}}$ is the classical free energy of the system constrained at the crossing seam, defined as
\begin{equation}
    \eu{-\beta F_\text{c}^{\text{cl}}}
    = \frac{1}{(2\pi\hbar)^{D}} \iint \eu{-\beta H_{\text{cl}}^{(0)}}\delta[\beta(V_0(\mathsf{x})-V_1(\mathsf{x}))] \,\text{d}\mathsf{x}\,\text{d}\mathsf{p},
    \label{Fc_cl}
\end{equation}
where $H_{\text{cl}}^{(0)}$ is the classical Hamiltonian of the reactant diabatic state, which is defined like $H_\text{RP}^{(0)}$ with $N=1$.

There are thus conceptual differences between Wolynes theory and GR-QTST\@.
Wolynes theory relies on a steepest-decent approximation to the time integral of the flux-flux correlation function in the golden-rule limit, \cite{wolynes}
whereas GR-QTST incorporates the physical requirement of energy conservation enforced by the virial energy estimator. \cite{GRQTST}
Both methods are approximations to the true quantum rate,
but can be shown to be very accurate for simple systems such as the spin-boson model. \cite{GRQTST}
It can also be shown that GR-QTST reduces to the classical rate, Eqn.~\eqref{kcl}, in the high-temperature limit of any system
when the ring polymers collapse. \cite{GRQTST}
However, the same is not necessarily true of Wolynes theory. \cite{nonoscillatory}
In particular,
we have shown that Wolynes theory can break down for systems with two or more different transition states,
due to the fact that only one $\lambda^*$ value is used which cannot simultaneously be optimal for all transition states.
In these cases at least, GR-QTST is expected to be more accurate as its rate is approximately independent of the choice of $\lambda^*$ and employs the energy constraint to ensure the correct sampling of each transition state. \cite{GRQTST2}

\section{Computational Methods} \label{compdet}

\begin{figure}
    \centering
    \includegraphics[width=.45\textwidth]{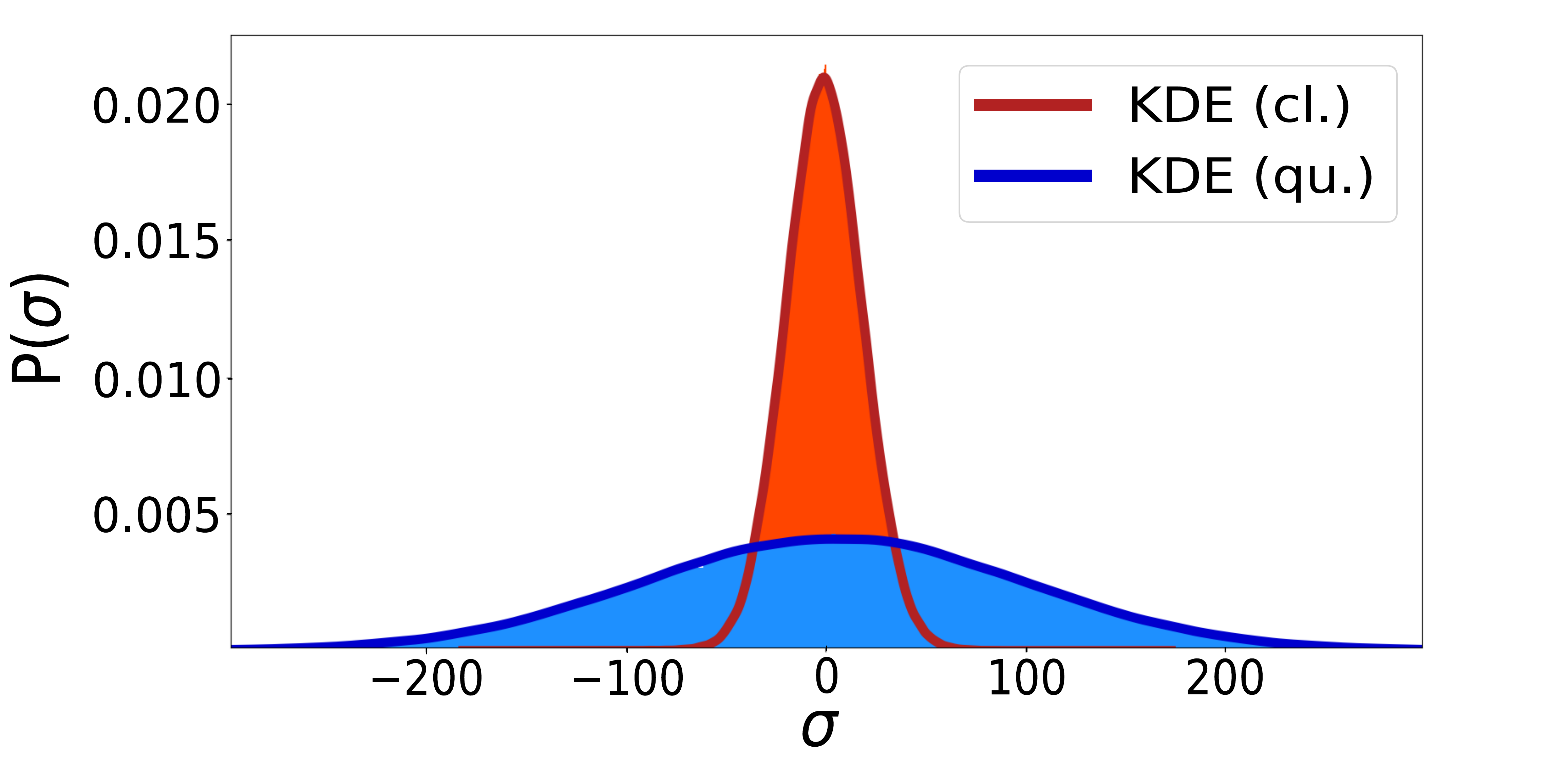}
    \caption{Histograms of the values of $\sigma=\sigma_{\lambda^{\ast}}\!(\mathbf{x})$
    sampled in an unconstrained simulation and the corresponding kernel density estimation (KDE) \cite{KDE_Book} in the classical and quantum limit. 
    The KDE at $\sigma = 0$ is used to obtain the constrained free energy, $F_\text{c}(\lambda^*)$, used in the GR-QTST method.
    }
    \label{histogram}
\end{figure}

We computed rates from Wolynes theory and GR-QTST in the classical limit according to Eqns.~\eqref{Wolynes_R} and \eqref{grqtst}. 
The free-energy term in the Wolynes rate was calculated 
using thermodynamic integration (TI) along the order parameter $\lambda$, 
\begin{equation}
    F_\text{u}(\lambda^{\ast})-F_0 = \int_0^{\lambda^{\ast}}\frac{\text{d}F_{\text{u}}}{\text{d}\lambda} \, \text{d}\lambda,
    \label{int_Fu}
\end{equation}
where the free-energy derivative
\begin{equation}
    \frac{\text{d}F_{\text{u}}}{\text{d}\lambda}
     = \left\langle V_1\left(\mathsf{x}^{(N_0)}\right) - V_0\left(\mathsf{x}^{(N_0)}\right) \right\rangle^{(\lambda)}
\end{equation}
can be obtained from sampling an unconstrained ring-polymer ensemble with the Hamiltonian $H_{\text{RP}}^{(\lambda)}$. \cite{Lawrence2018Wolynes,GRQTST2}
The constrained free energy, $F_\text{c}(\lambda^{\ast})$, was obtained by sampling from the same unconstrained ring-polymer ensemble with $H_{\text{RP}}^{(\lambda^{\ast})}$ and histogramming the probabilities of sampling a specific value of the function $\sigma_{\lambda^{\ast}}\!(\mathbf{x})$, which are defined by
\begin{equation}
\label{P}
P(\sigma) = \frac{ \iint \eu{-\beta_N H_{\text{RP}}^{(\lambda^{\ast})}} \delta \left( \sigma_{\lambda^{\ast}}\!(\mathbf{x}) - \sigma\right) \rmd \mathbf{x} \, \rmd \mathbf{p}}{\iint \eu{-\beta_N H_{\text{RP}}^{(\lambda^{\ast})}} \rmd \mathbf{x} \, \rmd \mathbf{p}}.
\end{equation}
The constrained free energy can then be expressed in terms of the sampling probability as
\begin{equation}
    \label{Fc}
    F_{\text{c}}(\lambda^{\ast}) = F_\text{u}(\lambda^{\ast}) -\frac{1}{\beta}\ln P(0).
\end{equation}
This procedure is computationally feasible if the unconstrained simulation samples
enough configurations which obey the constraint ($\sigma_{\lambda^{\ast}}\!(\textbf{x})$ = 0).
If this condition is not fulfilled, the $\delta$-TI method as described in Ref.~\citenum{GRQTST2} could be applied in combination to calculate $F_{\text{c}}(\lambda^{\ast})$.
However, this was not necessary for the system studied in this work as can be seen from the histograms in both the classical and the quantum limit shown in Fig.~\ref{histogram},
which are peaked around $\sigma = 0$.
Due to this connection in methodology, 
one can use unconstrained PIMD simulations as a first step towards 
either Wolynes theory or GR-QTST rate calculations.

Both the Wolynes and GR-QTST rates can also be evaluated in the classical limit.
The rate expressions  $k_{\text{Wolynes}}^{\text{cl}}$ and $k_{\text{GR-QTST}}^{\text{cl}}$ are very similar to Eqns.~\eqref{Wolynes_R} and \eqref{grqtst}, with the only difference being that the employed free energies are replaced by their classical counterparts $F_\text{u}^{\text{cl}}$,  $F_\text{c}^{\text{cl}}$ and $F_0^{\text{cl}}$. 
In the case of the classical Wolynes rate, the corresponding free-energy difference, $F_\text{u}^{\text{cl}} - F_0^{\text{cl}}$, can be computed analogously to the quantum case (Eqn.~\eqref{int_Fu}) with the ring polymer collapsed onto a single classical particle defined by the Hamiltonian
\begin{equation}
	\label{Hcl}
    H_{\text{cl}}^{(\lambda)} = \sum_{j=1}^{D}\frac{[p_j]^2}{2m_j} + (1-\lambda)V_0(\mathsf{x}) + \lambda V_1(\mathsf{x}).
\end{equation}
Alternatively, $F_\text{u}^{\text{cl}} - F_0^{\text{cl}}$ can be calculated by scaling up the mass of all the atoms (by multiplying $m_j$ by $\mu$ and taking the limit $\mu\rightarrow\infty$), which in effect collapses the ring polymer, making it behave classically. \cite{Lilienfeld_2011,massTIvarTransform,Wei_2016,doi:10.1021/acs.jctc.9b00596}
We found that both methods give results within each other's error bars for this system. 

Note that $k_{\text{Wolynes}}^{\text{cl}}$ is not the same as the classical rate expression in Eqn.~\eqref{kcl}, because Wolynes theory does not necessarily tend to the correct classical limit for general systems.
However, when $\frac{\text{d}F_\text{u}^{\text{cl}}}{\text{d}\lambda}$ is a linear function of $\lambda$ (which is the case for spin-boson models), one can show that $k_{\text{Wolynes}}^{\text{cl}}$ is the same as the rate of Marcus theory (Eqn.~\eqref{Marcus_R}), by plugging $\frac{\text{d}F_\text{u}^{\text{cl}}}{\text{d}\lambda}=\Lambda(1-2\lambda)$ into Eqn.~\eqref{Wolynes_R}.
In contrast,
as we have shown in Ref.~\citenum{GRQTST}, GR-QTST always reduces to the correct classical expression (\eqn{kcl})
in the high-temperature limit where the ring polymer collapses. 
This gives us two other possible methods for calculating the classical rate,
either by 
sampling the ensemble from the classical Hamiltonian $H_{\text{cl}}^{(\lambda^{\ast})}$ to obtain $F_{\text{c}}^{\text{cl}}$ following the same procedure given by Eqn.~\eqref{Fc},
or alternatively by scaling up the mass of all the atoms in a GR-QTST simulation.

\begin{figure}
        \centering
        \includegraphics[width=1\columnwidth]{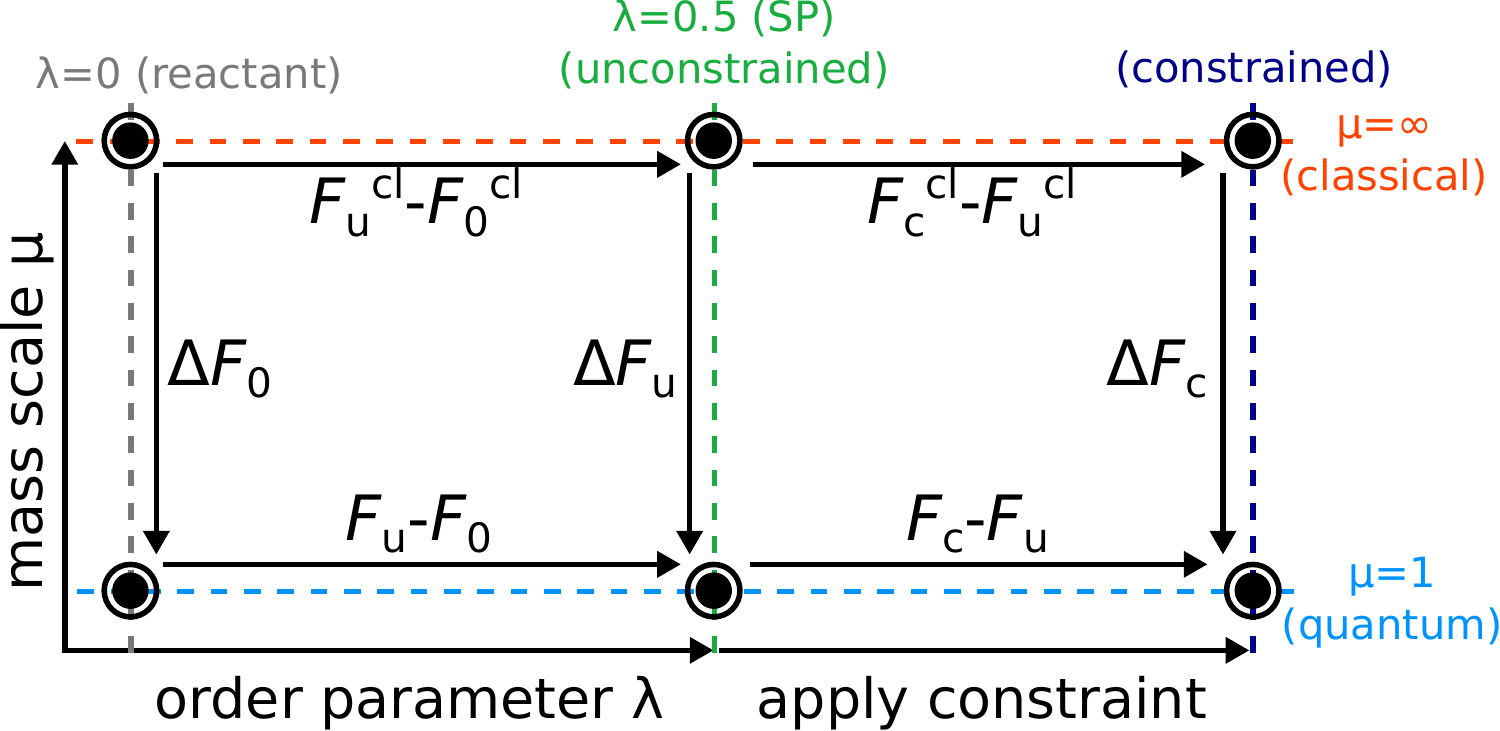}
        \caption{Thermodynamic cycle containing all the free energy integration schemes of Wolynes theory and GR-QTST in the quantum and classical limit.
        Constrained free energies are obtained at a value of $\lambda = \lambda^{\ast}$, which is the stationary point (SP) along the order parameter $\lambda$ of the unconstrained free energy $F_\text{u}^{(\lambda)}$.}
        \label{empty_cycle}
\end{figure}

In order to understand and to visualise the relation between all the different free energy terms in the rate theories introduced above, we constructed a thermodynamic cycle as shown in Fig.~\ref{empty_cycle}.

The free-energy calculations necessary to compute Wolynes theory
form the top and bottom horizontal thermodynamic paths of the left side of the cycle
and relate the reactant ($\lambda = 0 $) to the stationary point ($\lambda=\lambda^{\ast} = 0.5 $) in both the quantum and the classical limit. 
In order to compute GR-QTST, one additionally needs 
the free-energy calculations corresponding to the horizontal thermodynamic paths on the right side of the cycle. 
They can be viewed as an extension to the Wolynes free-energy calculations.

Each vertical thermodynamic path in the cycle represents a thermodynamic integration from the quantum nuclei to the classical limit 
using the mass-scaling factor $\mu$ as the order parameter (mass-TI). \cite{Lilienfeld_2011,massTIvarTransform,Wei_2016,doi:10.1021/acs.jctc.9b00596}
This means that each thermodynamic path of the cycle on the left can be calculated from \textit{independent} simulations.
Therefore, we use the left thermodynamic cycle to validate the accuracy and reliability of the free energies obtained from classical and quantum simulations, giving us further confidence in the Wolynes rates we computed.
The free-energy differences, 
 $\Delta F_0$ and $\Delta F_\text{u}$ were calculated 
by performing two sets of PIMD simulations with $H_{\text{RP}}^{(\lambda)}$ at $\lambda=0$ and $\lambda=\lambda^{\ast}$.  
The thermodynamic integrand was obtained for 10 different mass-scale factors from $\mu=1$ up to $\mu = 100$.
The contributions from larger $\mu$ values are also accounted for via a coordinate transform in the thermodynamic integration. \cite{massTIvarTransform,Wei_2016}
We do not compute $\Delta F_{\text{c}}$ as we found it to be numerically unstable to calculate due to the fact that the virial kinetic-energy estimator is not valid for constrained PIMD simulations.
This free-energy change can however be inferred by completing the cycle.

The ion--ion distance was treated using a fixed-atom implementation at an interionic distance of r = 5.5~\AA, which was determined to be the most probable interatomic distance for electron-transfer reactions. \cite{Kuharski,Bader}
The interactions in the aqueous ferrous--ferric system are defined by the interatomic forces and pseudopotentials described in Ref.~\citenum{Kuharski} with the exception of the water model.
In contrast to the formerly used rigid single point-charge (SPC) water model, \cite{Bader, Kuharski, spc-water}
we apply the flexible q-TIP4P/F water model, \cite{q-tip4p-f} which was specifically developed to suit PIMD simulations.
In particular, it can correctly capture the delicate balance between the competing quantum effects in water, compared to rigid or harmonic water models. \cite{q-tip4p-f,Ceriotti2016water}
Both, the q-TIP4P/F and SPC water models include electronic polarisation effects in a mean-field way \cite{polarisation2, polarisation3} and hence belong to the class of non-polarisable water models, which are computationally affordable and allow for extensive simulations. 
The application of an explicitly polarisable water model is crucial to describe effects in surface chemistry and clusters. \cite{polarisation1, jochen1}
The explicit treatment of polarisation is expected to lower the estimate of the reorganisation energy also in the aqueous ferrous--ferric system \cite{jochen1, jochen-RuRu, polarisation4} and it is therefore not without controversy to employ a non-polarisable water model. 
Ultimately, the choice of water model will of course affect the quantitative results, but will not hamper our ability to compare the different quantum rate theories. 
Nevertheless, it should be pointed out that several suggestions on polarisable water models and improved treatment of the solvent models were made in the literature. \cite{Song1993quantum2, polarisation1, polarisation2, polarisation3, jochen1, Blumberger2006RuRu}

The reorganisation energy of the ferrous--ferric system calculated with the q-TIP4P/F water model is
$108.5\pm0.9$ mHartree (68 kcal~mol$^{-1}$, 2.95 eV), 
which can be
compared to the 128~mHartree (80~kcal~mol$^{-1}$, 3.5 eV) value found in the previous work \cite{Bader, Kuharski} using a rigid water model.
Note that our setup gives a reorganisation energy only slightly closer to the experimental estimate of 2.1 eV. \cite{exp_reorg}

In order to obtain the rates for Wolynes theory and GR-QTST, we performed a set of PIMD simulations
using $N=24$ ring-polymer beads
at 13 values of the order parameter $\lambda$ (see Eqn.~\eqref{lambda}) for each integral $N_0$ value in the range $N_0 \in [0,12]$ to perform a thermodynamic integration along the order parameter $\lambda$.
In each case, we averaged over 10 starting configurations
picked randomly from a long MD simulation of the system with 265 water molecules in a cubic box of box length 20~\AA\ (to give a water density of 10$^{3}$~kg~m$^{-3}$) using periodic boundary conditions. 
The temperature was set to 300~K and kept constant using the Andersen thermostat.
Each simulation was then run 
under these conditions for 44,000 steps (including 4,000 steps of equilibration) with a timestep of 0.5 fs.
The only additional information required to obtain a GR-QTST rate from such an unconstrained PIMD simulation is a histogram of the sampled values of the energy constraint function $\sigma_{\lambda^{\ast}}\!(\mathbf{x})$.
This has a very minor computational cost as it
requires only one extra evaluation of the potential and forces on top of the $N$ which are performed anyway at each step of the MD simulation.
As we have chosen this setup in close analogy to the setup of Ref.\ \citenum{Kuharski}, the same considerations in terms of finite size effects and potential cutoffs apply.
All of the classical MD forces are calculated using \textsc{lammps}. \cite{lammps}

\section{Results and Discussion} \label{results}

\begin{figure}
    \centering
    \includegraphics[width=1\columnwidth]{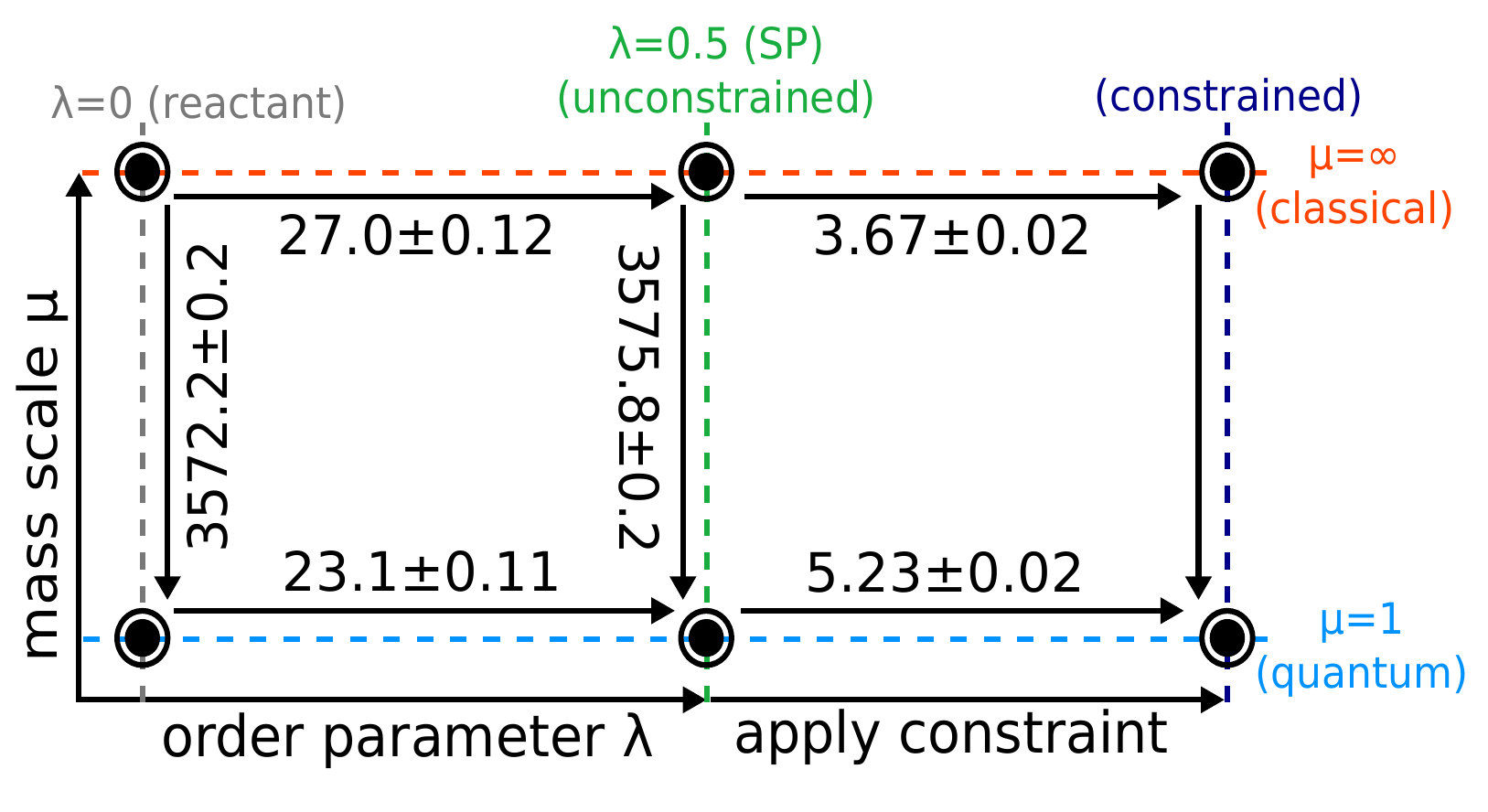}
    \caption{
    Thermodynamic cycles used to compute Wolynes theory and GR-QTST in the quantum and classical limit. Free energies are given in mHartree and error bars are
    of one standard deviation (1-sigma) calculated using block averaging \cite{Allen_Tildesley} and the error propagation formula.
    Cycle closure is observed for the unconstrained ensembles. The classical free energies were calculated using the mass-TI method. The classical Wolynes-theory calculation (thermodynamic integration along $\lambda$) performed using the collapsed ring-polymer method yields a free-energy change $F_{\text{u}}^{\text{cl}} - F_{0}^{\text{cl}} = 27.1 \pm 0.1$~mHartree.
    }
    \label{full_cycle}
\end{figure}

Our aim is to quantify the quantum effects present in the aqueous ferrous--ferric system and thereby to address the controversy of the magnitude of the effect of quantum tunnelling on the reaction rate.
In this section we present results from both Wolynes theory and our newly developed GR-QTST \cite{GRQTST,GRQTST2} 
and discuss the predictions for rate constants and isotope effects from these two different approaches.
We investigate possible pitfalls of each theory 
and discuss how they affect the rate of this system.
We then revisit the earlier studies \cite{Kuharski,Bader} to discover the effect of the improved water model 
and finally we compare our results with other quantum correction factors presented for the aqueous ferrous--ferric system in the literature. \cite{Bader}
Underlying all the rate calculations are free-energy differences which were defined in the thermodynamic cycle introduced in Section \ref{compdet}.
In Fig.~\ref{full_cycle} the results of these free-energy differences obtained 
from our simulations are given.

Before comparing the results obtained from the two quantum rate theories,
we first 
check for consistency of the calculations
of the various free-energy paths presented in Fig.~\ref{full_cycle}.
Note that the Wolynes rate is defined in terms of the quantities belonging to the thermodynamic cycle on the left, whereas GR-QTST depends also on those on the right.
As can be seen from Fig.~\ref{empty_cycle}, the free-energy change $\Delta \Delta F_\text{u}$ is defined in two alternative ways
\begin{subequations}
\begin{align}
    \Delta \Delta F_{\text{u}} &= (F^{\text{cl}}_{\text{u}}(\lambda^{\ast}) - F^{\text{cl}}_0) - (F_{\text{u}}(\lambda^{\ast}) - F_0 ) \label{g1} \\
    &= \Delta F_{\text{u}} - \Delta F_0 \label{g1b}.
    \end{align}
\end{subequations}
This free-energy change contributes exponentially to the quantum correction factor $\Gamma_\text{Wolynes}$ introduced later (see Section \ref{gamma_comp}, Eqn.~\eqref{our_gamma}.)
It can thus be obtained either as a difference of a thermodynamic integration in the quantum and the classical limit as described in Eqn.~\eqref{g1} and amounts to $3.9 \pm 0.2$~mHartree, or the difference between the mass-TI calculations at the reactant and stationary-point ensembles as defined in Eqn.~\eqref{g1b}, which gives $3.6 \pm 0.3$~mHartree.
The free-energy differences $\Delta F_0$ and $\Delta F_{\text{u}}$ of Eqn.~\eqref{g1b} obtained from the different mass-TI calculations are significantly larger in magnitude, because they include the change from classical to quantum nuclei and therefore include the zero-point energy of the system.
Further calculations are therefore made using the free energy differences as given by Eqn.~\eqref{g1} in order to avoid the numerical errors inherent to a subtraction of large numbers. 
Nevertheless, the consistency (within error bars) of the free-energy difference $\Delta \Delta F_{\text{u}}$ calculated via the two alternative routes also means that the left cycle is closed, which confirms that our Wolynes-theory simulations are converged.

Rates are then calculated according to Eqns.~\eqref{Wolynes_R}, \eqref{grqtst} and \eqref{kcl} from the changes in free energy and are listed in Table \ref{rates} for both the classical limit ($\mu \rightarrow \infty$) and the quantum limit ($\mu = 1$).
It is interesting to note that the quantum rate predictions of GR-QTST and Wolynes theory do not agree.
Both methods have been tested on the spin-boson model and give excellent and practically identical predictions of the quantum rate. \cite{GRQTST}
However, due to the conceptual difference of the two theories, in more complex systems
one cannot generally expect Wolynes theory and GR-QTST to predict similar rates.
This therefore implies that the aqueous ferrous--ferric electron-transfer reaction
is fundamentally more complex than the spin-boson model.
A second obvious conclusion is that 
since the two theories do not agree, at least one rate prediction must be inaccurate.
In the following we analyse the two methods to discuss these points.

\begin{table}
\caption{Calculated rates in atomic units in the classical and quantum limit using different rate theories. 
An alternative calculation of the classical Wolynes rate using a collapsed ring polymer gives an almost identical rate ($7.2\pm1.0\times10^{-11}$).
}
\label{rates}
\resizebox{\columnwidth}{!}{
\begin{tabular}{l|lll}
rate                 & classical & quantum (H$_2$O) & quantum (D$_2$O) \\ 
\hline
$k_{\text{Marcus}}$  & $7.0\pm1.7\times10^{-11}$   & -                           & -  \\
$k_{\text{cl}}$      & $7.3\pm1.0\times10^{-11}$   & -                           & - \\
$k_{\text{Wolynes}}$ & ${7.6\pm1.0\times10^{-11}}$ & $5.3\pm0.7\times10^{-9}$    & $2.7\pm0.3\times10^{-9}$ \\
$k_{\text{GR-QTST}}$ & $6.3 \pm 0.9\times10^{-11}$ & $7.6\pm0.9\times10^{-10}$   & $4.5\pm0.4\times10^{-10}$  \\
\end{tabular}
}
\end{table}

\subsection{Making or breaking of Wolynes theory}

As presented in Ref.~\onlinecite{GRQTST2}, a break-down of Wolynes theory 
occurs when the system under investigation exhibits multiple distinct transition states,
which can lead to an overprediction of the rate by orders of magnitude.
There are a number of criteria that serve as indicators to identify whether Wolynes theory is applicable,
although the absence of these features does not exclude the possibility of at least a minor break down of Wolynes theory.
The first criterion is whether the rate tends to the correct classical limit as the masses are scaled up.
Second, one should investigate the ensemble of paths sampled by the unconstrained simulation
to check that these are centred around energy-conserving paths like the instantons.
Finally, any evidence for the existence of multiple transition states
with a range of different $\lambda$ values 
would suggest that Wolynes theory is not valid as it cannot simultaneously satisfy the condition for each transition state.

\begin{figure}
    \centering
    \includegraphics[width=.4\textwidth]{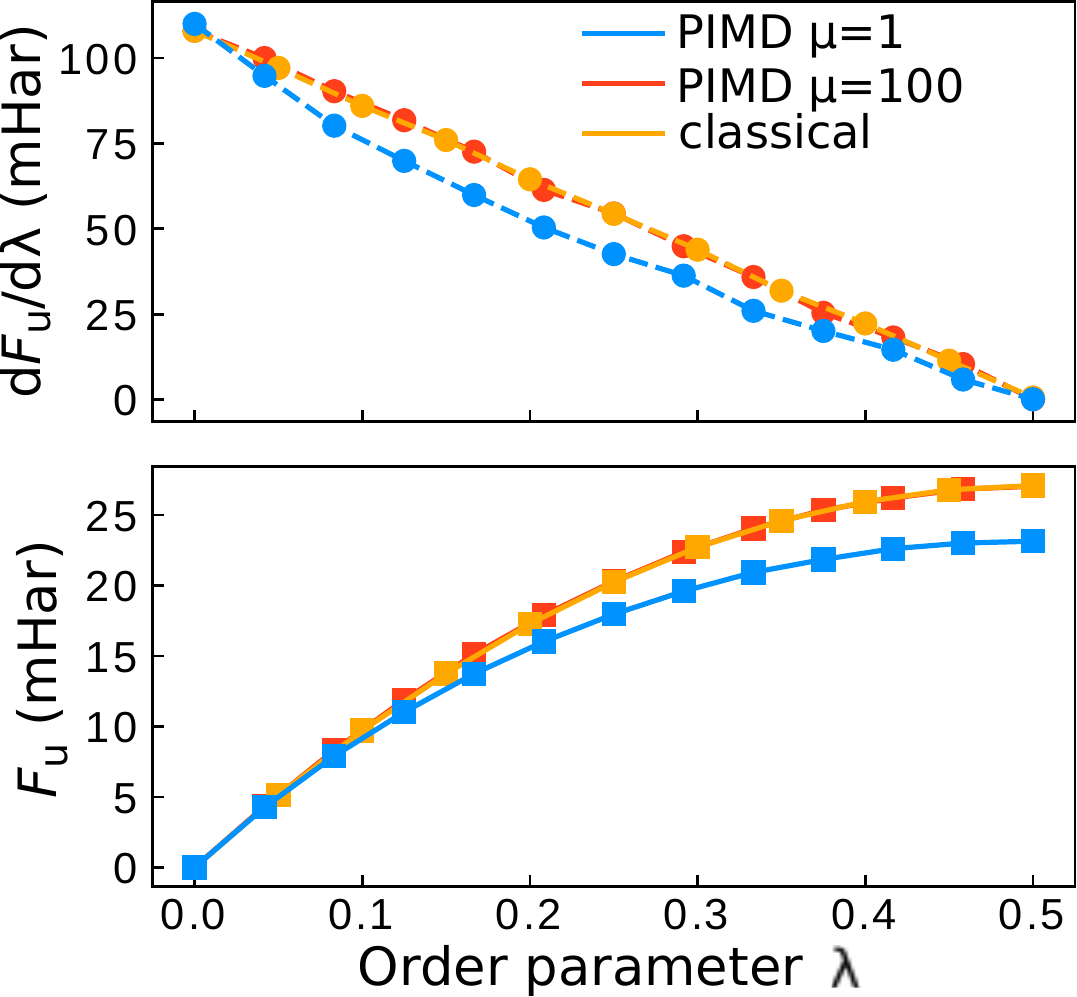}
    \caption{Plot of the Wolynes free energy derivative and free energy.
    The ``classical'' result refers to the simulation with a collapsed ring polymer (Eqn.~\eqref{Hcl}).
    The error bars are smaller than the symbol size.
    }
    \label{Wolynes_F}
\end{figure}

In the classical (high-temperature or heavy mass) limit, in contrast to Wolynes theory, GR-QTST is known to tend to the correct classical rate, \cite{GRQTST}
and our simulations are in agreement with this (see Table \ref{rates}).
As suggested above, there is no rigorous argument that requires Wolynes theory to correctly predict the true classical rate and this thus provides a good check that the method gives physically sensible results.
In this case, it does give the correct result within the error bars.
This success of Wolynes theory in the classical limit can be related to the approximately linear behaviour of the free-energy derivative with respect to the order parameter, $\lambda$, as shown in Fig.~\ref{Wolynes_F}.
This is a clear indication that the linear-response approximation is valid for this system in the classical limit,
as was already discussed in previous work. \cite{Kuharski, Song1993quantum2, ChandlerET}
As a consequence, the classical limit of the aqueous ferrous--ferric system strongly resembles a spin-boson model, where Wolynes theory is known to perform well. 
The same also explains the good agreement of Marcus theory with the exact classical rate, because Marcus theory, similarly to Wolynes theory, is known to perform well for this model. 
For a quantitative comparison, we note that the free-energy barrier according to Marcus theory is $\Lambda/4=27.1$~mHartree, which is in excellent agreement with that found from Wolynes theory in the classical limit ($27\pm0.12$~mHartree).
Note that this is not always the case for asymmetric reactions, where the linear-response approximation is commonly seen to break down. \cite{Blumberger2006Ag}
The aqueous ferrous--ferric system therefore does not exhibit a break-down of Wolynes theory in the classical limit, in contrast to the model systems tested in Ref.~\onlinecite{GRQTST2}.

The investigation of the second qualitative indicator of break-down of Wolynes theory is the distribution of values of $\sigma=\sigma_{\lambda^{\ast}}\!(\mathbf{x})$ which are sampled in the unconstrained ensemble.
If this distribution had a negligible population at $\sigma=0$,
which was the case for the system under study in Ref.~\onlinecite{GRQTST2},
it would imply that the paths being sampled have no connection to the energy-conserving instantons and would be a clear sign of the break-down of Wolynes theory.
However, as shown in Fig.~\ref{histogram}, Wolynes-theory calculations sample a uni-modal distribution peaked around $\sigma = 0$.
This therefore neither confirms nor disproves a break-down of Wolynes theory in this system.

A more detailed observation can be made from a comparison to another quantum rate theory, namely instanton theory.
Instanton theory is not rigorously applicable to reactions in solution \cite{InstReview} and we cannot therefore use it to calculate the rate.
However, we can nonetheless acquire a qualitative insight into the tunnelling pathways of the system by obtaining a set of optimised instanton paths \cite{GoldenRPI} on different solvent configurations randomly taken from a MD simulation (250 configurations).
All instanton optimisations were able to find non-trivial tunnelling pathways,
which suggests that nuclear tunnelling is a significant contributor to the quantum rate enhancement.
The water molecules beyond a radius of 5~\AA\ of either Fe ion were fixed and only water molecules within this circumference were optimised in the instanton calculations (approximately 36 flexible water molecules). 
\begin{figure}
    \centering
    \includegraphics[width=0.55\columnwidth]{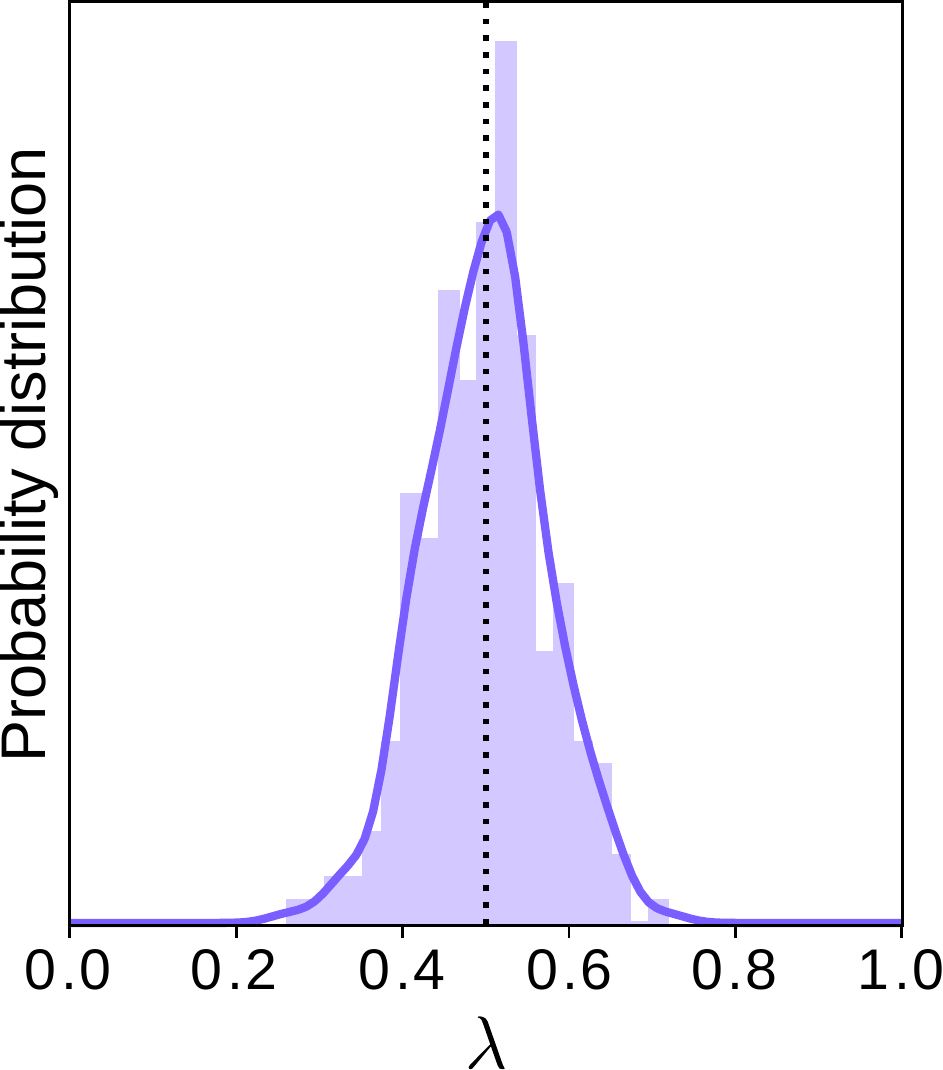}
    \caption{ 
    Distribution of the $\lambda$ values found from the ensemble of instantons.
    The dotted line shows $\lambda^{\ast}=0.5$, which is the value appropriate for Wolynes theory,
    and due to symmetry is also the average of the ensemble of instantons.
    }
    \label{inst_dist}
\end{figure}
Analysing the ensemble of instantons clearly shows that the aqueous ferrous--ferric system has multiple transition states.
As shown in Fig.~\ref{inst_dist}, the instantons have a range of different order parameters $\lambda$,
which are found by a stationary-action principle. \cite{Perspective}
A distribution of instantons occurs because even though the system is globally symmetric, it is locally asymmetric around each instanton and this is another sign that the system differs from the spin-boson model, which has only one instanton with $\lambda=0.5$.
The order parameters for each instanton cannot all be simultaneously satisfied by the single choice made by Wolynes theory of $\lambda^*=0.5$,
which suggests that Wolynes theory may break down for this system and overpredict the rate in the quantum limit.
However, a broad uni-modal distribution as observed here is clearly a much safer scenario for Wolynes theory than the system tested in Ref.~\citenum{GRQTST2} for which this equivalent plot would have two peaks on either side of $\lambda^*$.
The fact that in this case the distribution is uni-modal and centred at $\lambda=0.5$ (with a standard deviation of 0.07) suggests that the break down will be less severe.

\subsection{Discussion of the GR-QTST result}
\label{gamma_comp}
GR-QTST, in contrast to Wolynes theory, can treat systems with multiple transition states correctly, \cite{GRQTST2} and similarly to Wolynes theory, gives an accurate result for the spin-boson model in both the classical and the quantum limit. \cite{GRQTST}
In fact, the close agreement of GR-QTST and Wolynes theory for a spin-boson model \cite{GRQTST} is another argument towards the aqueous ferrous--ferric system being badly approximated by a spin-boson model in the quantum limit, because for this system the two rate theories disagree (see Table \ref{rates}).

\begin{figure}
    \centering
    \includegraphics[width=.8\columnwidth]{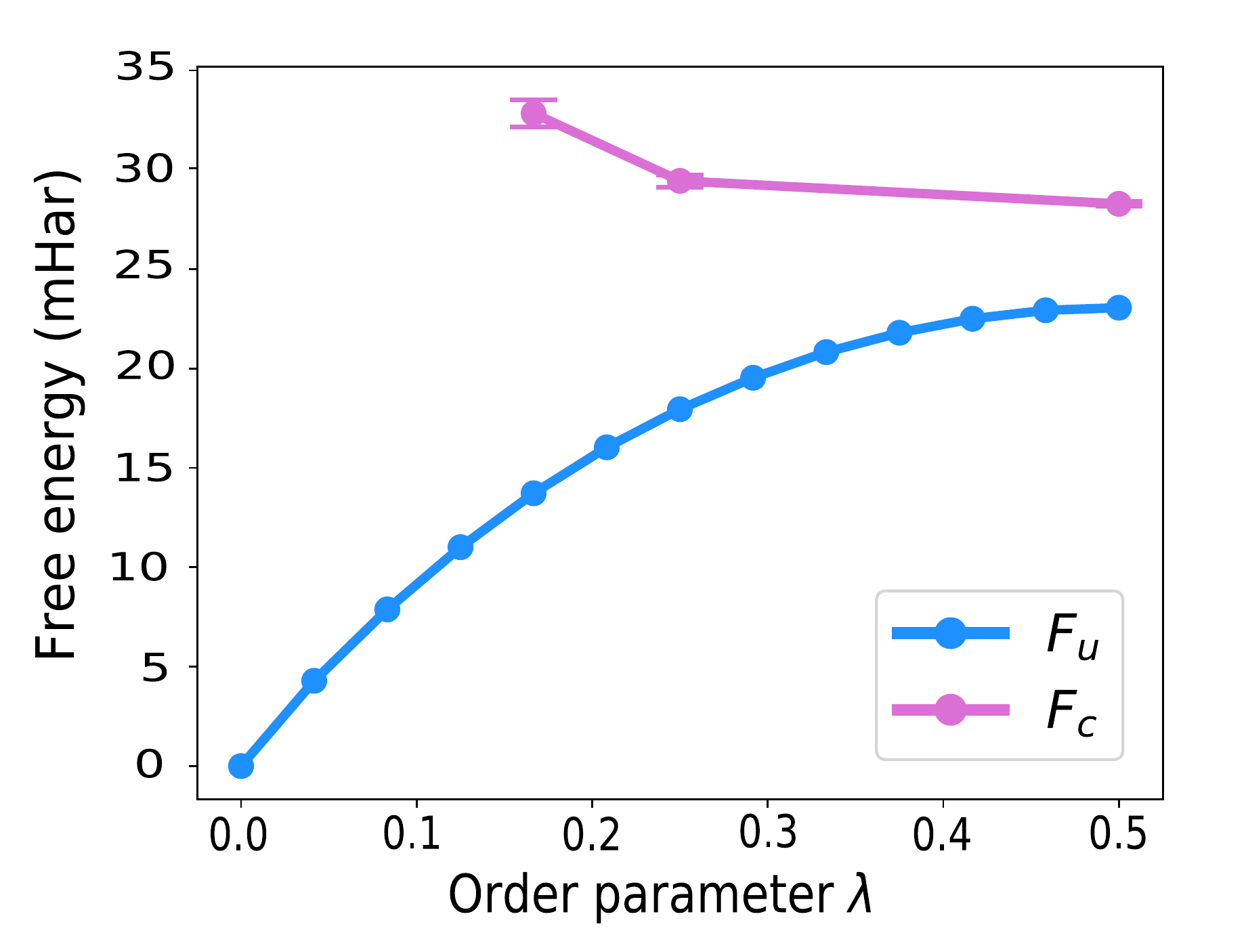}
    \caption{Comparison of the curvature of the free energies of Wolynes (unconstrained) and  GR-QTST (constrained) along the path-splitting parameter $\lambda$.
    We obtained the constrained free energies $F_{\text{c}}(\lambda)$ at values of $\lambda\ne0.5$ by combination of the $\delta$-TI method and the histogramming. 
    The error bars on the unconstrained calculations are smaller than the marker size.}
    \label{curv}
\end{figure}

Next to the discrepancy in rates, the curvature of the constrained free energy $F_\text{c}(\lambda)$
shown in Fig.~\ref{curv} 
can be utilised to argue against the applicability of linear-response theory and therefore the approximation of this system by a spin-boson model.
Unlike the observation for a spin-boson model, for the atomistic system under study the constrained free energy $F_{\text{c}}(\lambda)$ is curved upwards in extreme regions of the order parameter $\lambda$, i.e.\ far from the optimal order parameter $\lambda^{\ast}$. 
In the earlier investigation of GR-QTST on spin-boson models, \cite{GRQTST} we found that the constrained free energy $F_\text{c}(\lambda)$ curves down when going away from the optimal order parameter $\lambda^{\ast}$.
This curvature behaviour becomes more prominent with an increased number of degrees of freedom and is already significant for 8 degrees of freedom,
which can be understood based on the analysis presented in Ref.~\onlinecite{GRQTST}.
The fact that the curvature behaviour differs from this is a further indicator for the aqueous ferrous--ferric system not being well described by a spin-boson model.

The curvature of the constrained free energy $F_\text{c}(\lambda)$ is also of interest
as it is an indication of the size-consistency error of GR-QTST\@.
In Ref.~\onlinecite{GRQTST} we showed that as more degrees of freedom are added to the system,
the plot of $F_\text{c}(\lambda)$ becomes more and more curved.
However, we also argued that no matter how large the system, it becomes flat in the classical limit,
and that for a spin-boson model, the value of $F_\text{c}(\lambda^*)$ remains stable even in the quantum case.
There is however the possibility that this could lead to an error for more complex systems such as the aqueous ferrous--ferric reaction studied here.
However, in the broad vicinity of the stationary point $\lambda^{\ast}$ the curvature of the constraint free energy is approximately flat. 
This is a good sign that there is no serious error being made by the GR-QTST method.

\subsection{New and old controversies of the aqueous ferrous--ferric electron transfer}

Of particular interest is the enhancement of the electron-transfer rate due to nuclear quantum effects,
which can be quantified by the quantum correction factor $\Gamma$,
defined as the ratio of the quantum and the classical rate.
Already 30 years ago the investigation of the aqueous ferrous--ferric system led to a broad range of predictions for this quantity, \cite{Bader, Song1993quantum2, Siders1981quantum}
and no conclusive argument 
could be made at the time
for which was correct. 
We reopen this controversy by adding the results of our new GR-QTST approach to the discussion.

We define $\Gamma$ for each quantum rate theory, for example
\begin{equation}
    \Gamma_{\text{Wolynes}} \equiv \frac{k_{\text{Wolynes}}}{k_{\text{Wolynes}}^{\text{cl}}} = \frac{A_{\text{Wolynes}}}{A_{\text{Wolynes}}^{\text{cl}}} \, \eu{-\beta \Delta \Delta F_{\text{u}}},
    \label{our_gamma}
\end{equation}
with the exponent $\Delta \Delta F_{\text{u}}$ defined in Eqn.~\eqref{g1}. 
We use $A_\text{Wolynes}$ as short-hand for the pre-exponential factor in Eqn.~\eqref{Wolynes_R}
and $A_\text{Wolynes}^\text{cl}$ is its classical counterpart.

The prefactor ratio $A_\text{Wolynes}/A^{\text{cl}}_\text{Wolynes}$ is generally not exactly equal to 1, although the equivalent term for GR-QTST is identically 1 and does not therefore appear.
Accordingly, the quantum correction factor for GR-QTST can be defined simply as $\Gamma_{\text{GR-QTST}}=\eu{-\beta\Delta \Delta F_{\text{c}}}$ with the exponent 
\begin{equation}
    \Delta \Delta F_{\text{c}}
    = \left (F_{\text{c}}(\lambda^{\ast}) - F_0 \right) - (F^{\text{cl}}_{\text{c}}(\lambda^{\ast}) - F^{\text{cl}}_0).
\end{equation}
Defining the quantum correction factor in this way gives the most fair comparison between methods as it would allow for some error cancellation in the case that Wolynes theory breaks down and overestimates the rate
in both the classical and quantum limit. $\Gamma$ therefore describes the quantum rate enhancement described by a given theory and avoids inconsistencies by cross-comparison of different theories.

The quantum correction factor gives a measure of nuclear tunnelling in the reaction.
Note that it is tricky to rigorously separate the tunnelling contribution from other NQEs such as vibrational quantization \cite{Schatz1987tunnel} and 
it therefore technically quantifies the rate enhancement from all NQEs.
In the case of the spin-boson model, however, the potentials are harmonic which ensures that zero-point energy is the same everywhere.
Tunnelling is thus the only factor contributing to the quantum rate enhancement in this case. \cite{Bader}
Although this argument does not hold rigorously for the atomistic system,
we assume that nuclear tunnelling continues to play an important role in this system
and present evidence in the SI to support this based on the instanton optimisations described above.

\begin{table}
\caption{Quantum correction factors as defined in Eqn.~\eqref{our_gamma} or accordingly. Note that the result reported in Ref. \citenum{Bader} only describes the exponential contribution to the rate. 
}
\label{tab_gamma}
\begin{tabular}{l|ll}
                          &     This work & Ref. \citenum{Bader} \\
\hline
$\Gamma_{\text{Wolynes}}$ & $70 \pm 13$       & $65 \pm 6 $               \\
$\Gamma_{\text{GR-QTST}}$ & $12 \pm 2$        & -                         \\
$\Gamma_{\text{spin-boson}}$  &  83   &   36 \\
\end{tabular}
\end{table}

The quantum correction factors obtained from various rate theories are 
reported in Table \ref{tab_gamma}. Our results from Wolynes theory are in agreement with similar calculations performed by Chandler and co-workers \cite{Bader} despite the fact that we employed a different water model.
This appears to be a bit of a coincidence because
when we map our system (with the flexible q-TIP4P/F water) on to a spin-boson model following the same procedure as in Ref.~\citenum{Bader} (for which the spectrum is shown in the SI)
and solve for the exact quantum rate, \cite{Weiss} we find a large quantum correction factor of $\Gamma_{\text{spin-boson}} = 83$.
This is more than a factor of two larger than the tunnelling enhancement ($\Gamma_{\text{spin-boson}}=36$) found for the SPC water model. \cite{Bader}
In part this deviation can be attributed to the contribution of high-frequency modes (H-bond bending and stretching) of the flexible water model.
Integration of the spectral density gives a reorganisation energy of 3.1$\pm0.2$ eV, in which 90\% comes from the low frequency modes.
It is, however, due to the NQEs of the high-frequency modes that 
we obtain a larger quantum correction factor, \cite{jochen-RuRu}  $\Gamma_{\text{spin-boson}}$, with q-TIP4P/F water, as can be shown from the fact that
if we only accounted for the low-frequency modes, we would predict that 
$\Gamma_{\text{spin-boson}}$ was only 15.
The reason that $\Gamma_{\text{spin-boson}}=36$ for SPC water is not as low as this
is a result of its overall larger reorganisation energy ($\Lambda = 3.5$ eV).
In order to show this
we used the low-frequency part of the q-TIP4P/F spectral density and scaled it up to produce $\Lambda=3.5$ eV.
In this case $\Gamma_{\text{spin-boson}}$ increases to 28, which is closer to the result of SPC water.
This analysis shows that the two water models are significantly different.
It so happens that, due to the two competing effects of flexibility and the lower reorganisation energy of the q-TIP4P/F model, we obtain similar a result within Wolynes theory,
but a different result for the spin-boson model.

It also appears to be a coincidence that the quantum correction factor predicted by Wolynes theory using the flexible q-TIP4P/F and the spin-boson model give such similar results.
We have presented a number of arguments throughout this paper to explain why one would not in general expect them to be the same, and indeed this was not found to be the case in the study by Chandler and co-workers. \cite{Bader}
The most important finding of this work is that the predictions from
Wolynes theory and GR-QTST differ significantly.
Each of the three methods presented in Table~\ref{tab_gamma} employs a different approximation
and it is difficult to determine which (if any) is correct as no exact quantum-mechanical rate for the aqueous ferrous--ferric system can be computed.
A comparison to experimental results is, however, another possible aspect to investigate.

For the ferrous--ferric system experimental isotope effects are available and the presence of a kinetic isotope effect is proof that NQEs play a role in this reaction as the classical rate does not depend on the masses of the atoms.
The experimental estimate of the isotope effect is in the range of $k^{\text{H}}_{\text{exp}}/k^{\text{D}}_{\text{exp}} = 1.7 - 2.0$, \cite{isotope-exp1, isotope-exp2} which compares well to the ratio $k^{\text{H}}_{\text{Wolynes}}/k^{\text{D}}_{\text{Wolynes}} = 2.0\pm0.4$ that we find by employing Wolynes rate theory  
for both isotopes.
Our Wolynes-theory calculations show an increase of the free energy difference from $(F_{\text{u}} - F_0)_{\text{H}} = 23.1 \pm 0.11$~mHartree for the hydrogen isotope to $(F_{\text{u}} - F_0)_{\text{D}} = 23.8 \pm 0.12$~mHartree for the deuterium isotope.
Our prediction of the isotope effect from the GR-QTST calculations is $k_{\text{GR-QTST}}^{\text{H}}/k_{\text{GR-QTST}}^{\text{D}} = 1.7 \pm 0.3$ and thus also lies within the range of experimental findings.
The GR-QTST prediction is slightly lower than that of Wolynes theory, because nuclear tunnelling plays a smaller role (see Table \ref{tab_gamma}).
Ref.~\citenum{Bader} reported an isotope factor of $k_{\text{Wolynes}}^{\text{H}}/k_{\text{Wolynes}}^{\text{D}} = 2.6 \pm 0.5$.

Although our main focus is on the comparison of different quantum rate theories,
in order to justify our comparison with the experimental isotope effect, we must also consider the accuracy of the atomistic model.
As we have already discussed earlier the choice of the water model has a crucial effect on the predicted rates and NQEs.
Marcus and co-workers \cite{Song1993quantum2} predict a significantly lower quantum correction factor of $\Gamma_\text{spin-boson}=9.6$
(using the spin-boson model \cite{Siders1981quantum} with an experimental spectral density) than Chandler and co-workers' results.
Note that this result cannot be taken as a benchmark as it is based on the spin-boson model, which we argue is a questionable approximation.
They proposed in Ref.~\citenum{Song1993quantum2} that the discrepancy in the two predictions 
might be due to the neglect of electronic polarisation and flexibility of water in Ref.~\citenum{Bader}, although as they are competing effects they may cancel to a certain extent. \cite{jochen1}
We have explicitly included the flexibility of water molecules in our study and found that it can have a significant impact on the rate enhancement due to nuclear tunnelling.
Nonetheless, although we predict a lower reorganisation energy (2.95~eV) using the flexible water model than Chandler and co-workers (3.5~eV)\cite{Bader} do with the rigid one,
neither water model reproduces the experimentally estimated reorganisation energy (2.1~eV). \cite{exp_reorg}
If we were to include electronic polarisation as well, one would expect the reorganisation energy 
to decrease.
This is turn could result in reduced tunnelling effects.\cite{jochen1}

This work therefore aims to reopen the discussion on the question of tunnelling enhancement in the atomistic model of the ferrous--ferric electron transfer as presented here.
The majority of previous studies on this question have concentrated on exploring the effect of changing the spectral density of the bath \cite{Song1993quantum2} or improving the description of the atomistic model. \cite{jochen-RuRu}
We have, however, explored how the predictions depend on the choice of quantum rate theory used for a given system Hamiltonian.
We have shown that a number of different methods obtain different rate predictions and currently no decisive argument for which (if any) gives the correct result can be made.
This is reminiscent of the controversy surrounding the quantum tunnelling effect in
the Azzouz--Borgis model of proton transfer, 
for which various approximate quantum rate theories do not agree.
\cite{HammesOnAzzouz-Borgis, RosanaOnAzzouz-Borgis, CraigOnAzzouz-Borgis, MUSTreview}
Any further study of these problems however provides valuable insights into the accuracy, applicability and pitfalls of the quantum methods applied.

\section{Conclusions}
\label{conclusion}

In this study, we have presented the first application of GR-QTST to an atomistic system of electron transfer and thereby obtained estimates for the reaction rate, isotope and tunnelling enhancement effects from this new method.
The aqueous ferrous--ferric electron-transfer reaction has been extensively studied in the past
and yet no conclusive quantitative answer for the contribution of nuclear tunnelling to the reaction rate has been given.
In fact the previously predicted quantum correction factors span an order of magnitude \cite{Kuharski, Bader,Siders1981quantum, Song1993quantum2} and our new prediction is at the lower end of this range.

All the methods tested reproduce the correct rate in the classical limit.
GR-QTST is guaranteed to do this for any system,
whereas Wolynes theory and Marcus theory are correct only if the linear-response approximation is valid.
This implies that the classical limit of this reaction can be adequately described by a spin-boson model.
This observation 
may lead one to believe that the spin-boson model is a valid approximation of the aqueous ferrous--ferric system also in the quantum limit. 
However, we could show the unsuitability of this assumption by comparing Wolynes theory \cite{wolynes} and our newly developed GR-QTST. \cite{GRQTST, GRQTST2}
In an earlier study it was observed that GR-QTST and Wolynes theory predict the same rate for a spin-boson model in the quantum limit. \cite{GRQTST}
In contrast to this, the two theories predict quantum correction factors that differ by a factor of~6 for the atomistic system, therefore making it impossible to argue that a spin-boson model is a good approximation in this case.

The obvious next question aims at resolving the disagreement of Wolynes theory and GR-QTST and in order to do so the possible pitfalls of each method were investigated.
We have explained that there is a risk that Wolynes theory may break down and overpredict the rate, especially in liquid systems like that under study. This may occur whenever a reaction contains multiple distinct transition states. \cite{GRQTST2}
We optimised a set of instantons in the system and found that 
their order parameters, $\lambda$, were distributed around $\lambda^*=0.5$.
This implies that there are indeed multiple transition states in this system,
although they are more similar to each other than the extreme cases studied in Ref.~\citenum{GRQTST2}.
The error 
made by Wolynes theory is thus expected to be less severe, but may still exist to some extent.
GR-QTST is also not without its flaws and we have shown in previous work that
the theory may suffer from size inconsistency.
In a model system the addition of many degrees of freedom leads to a strongly $\lambda$-dependent $F_\text{c}(\lambda)$ curve, 
which may degrade the rate predictions.
Nevertheless, for the high-dimensional aqueous ferrous--ferric system, we observe only a minor curving of the constrained free energy of GR-QTST far from the optimal order parameter $\lambda^{\ast}$,
which is not expected to significantly degrade the result.

Considering the discrepancy in the predicted tunnelling enhancement makes it however apparent that at least one theory must be inaccurate for this system.
A hypothesis that Wolynes theory is overpredicting the rate due to the multi-instanton nature of the system
cannot be excluded from the results of our calculations.
However, none of the studies provide us with a conclusive case 
to prove this statement,
nor to rule out the possibility of an error on the part of GR-QTST\@.
By revisiting the controversial question of the nuclear tunnelling effect in this system,
we could however 
show that the dynamics in this system deviates from those of the spin-boson model
and raises the question of the applicability of this model for simulating atomistic systems.
We hope to further elucidate the question of appropriateness of these quantum rate theories, by applying GR-QTST to a more complex atomistic system, where Wolynes theory may show more distinctive break-down behaviour.
    
\section*{Conflicts of interest}
There are no conflicts to declare.

\section*{Acknowledgements}
We thank Manish J. Thapa, Joseph E. Lawrence and Prof.\ Jochen Blumberger for the helpful discussions.
The authors acknowledge financial support from the Swiss National Science Foundation through Project 175696.

\providecommand*{\mcitethebibliography}{\thebibliography}
\csname @ifundefined\endcsname{endmcitethebibliography}
{\let\endmcitethebibliography\endthebibliography}{}

\end{document}


\hyphenation{ALPGEN}
\hyphenation{EVTGEN}
\hyphenation{PYTHIA}

\title{Supporting Information:
Revisiting nuclear tunnelling in the aqueous ferrous--ferric electron transfer.}

\author{Wei Fang}
\thanks{These authors contributed equally}
\author{Rhiannon A. Zarotiadis}
\thanks{These authors contributed equally}
\author{Jeremy O. Richardson}
\email{jeremy.richardson@phys.chem.ethz.ch}
\affiliation{Laboratory of Physical Chemistry, ETH Z\"urich, 8093 Z\"urich, Switzerland}

\date{\today}

\begin{abstract}
In this supporting information we provide further details of the simulations reported in the manuscript and additional discussions in support of the conclusions reached.
\end{abstract}

\maketitle

\section{Marcus theory }
\begin{figure}[!ht]
    \centering
	\includegraphics[width=.6\textwidth]{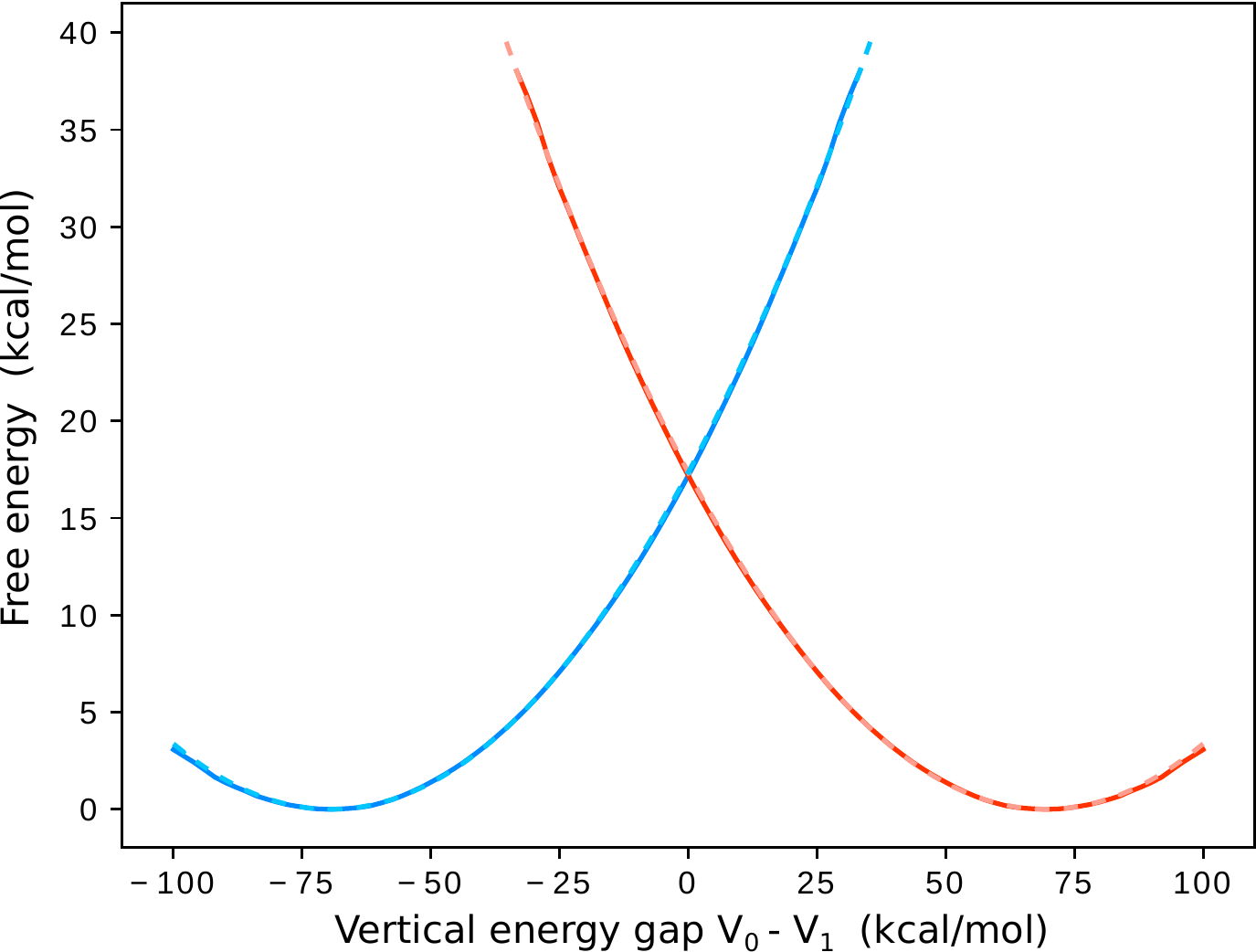}
    \caption{Classical free energy curves for the ferrous--ferric electron transfer with the flexible q-TIP4P/F water model, obtained via umbrella sampling (as in Ref.~\cite{Kuharski}) and the weighted histogram analysis method \cite{TuckermanBook}.
    The Marcus parabolas depicted using dashed lines are almost identical. 
    The free-energy barrier obtained from this calculation is 16.9~kcal~mol$^{-1}$.}
    \label{marcus_water}
\end{figure}

\section{Mass thermodynamic integration for Wolynes free energy}
\begin{figure}[H]
    \centering
    \includegraphics[width=.7\textwidth]{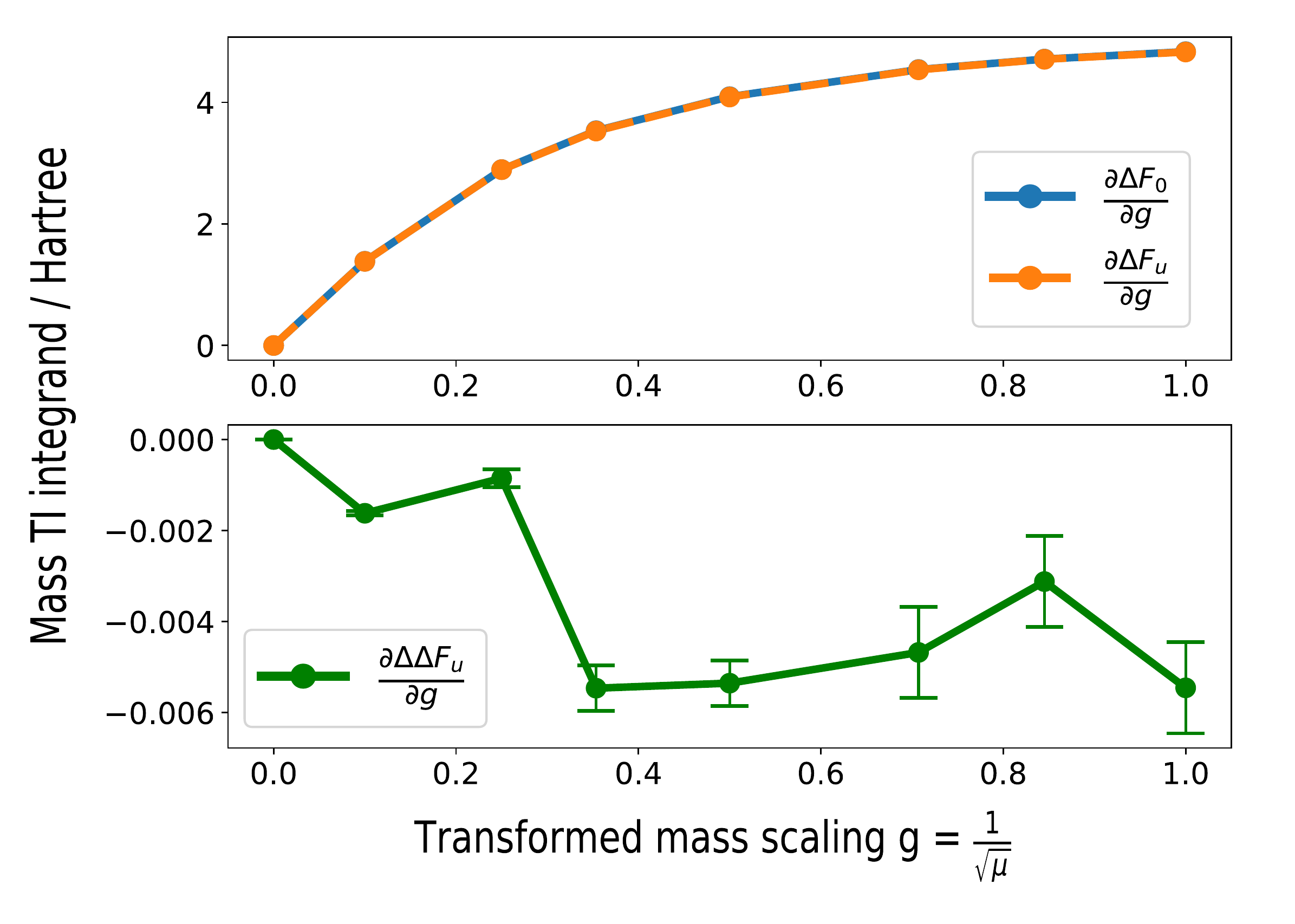}
    \caption{
    (a) Mass-TI integrands (formula see Ref.~\citenum{Wei_2016}) $\frac{\partial \Delta F_\text{0}}{\partial g}$ and $\frac{\partial \Delta F_\text{u}}{\partial g}$ obtained from path integral molecular dynamics simulations.
    $\Delta F_\text{0}$ and $\Delta F_\text{u}$ are defined in Eqn.~(17) in the main text. 
    $g = \sqrt{1/\mu}$ is the transformed mass scaling, meaning the masses of all atoms are scaled by $\mu$ in the corresponding simulation.
    (b) Difference of the mass-TI integrands $\frac{\partial \Delta \Delta F_\text{u}}{\partial g}=\frac{\partial \Delta F_\text{u}}{\partial g}-\frac{\partial \Delta F_\text{0}}{\partial g}$.
    }
    \label{Wolynes_large_F}
\end{figure}

\section{Instanton action}
In order gain qualitative insight into whether the quantum correction factor $\Gamma$ has tunnelling contributions or simply dominated by zero point energy contributions, we further analysed our golden-rule instanton \cite{Perspective,GoldenGreens} results.
The instanton rate expression can be formally expressed as \cite{Perspective,GoldenGreens},
\begin{equation}
    k_{\text{inst}}=A_{\text{inst}} \, \eu{-S/\hbar}.
\end{equation}
$S$ is the instanton action and $A_{\text{inst}}$ is a prefactor.
While the instanton rate theory is not rigorously applicable to reactions in solution \cite{InstReview}, the instanton action $S$ provides qualitative insight into tunnelling effects.
We can defined an approximate tunnelling factor for each instanton using
\begin{equation}
\label{gamma_inst}
    \Gamma_{\text{inst}} \approx \eu{-(S/\hbar-\beta V^{\ddagger})},
\end{equation}
in which $V^{\ddagger}$ as the classical transition state corresponding to the same reaction mechanism as the instanton.

\begin{figure}[H]
    \centering
    \includegraphics[width=0.33\columnwidth]{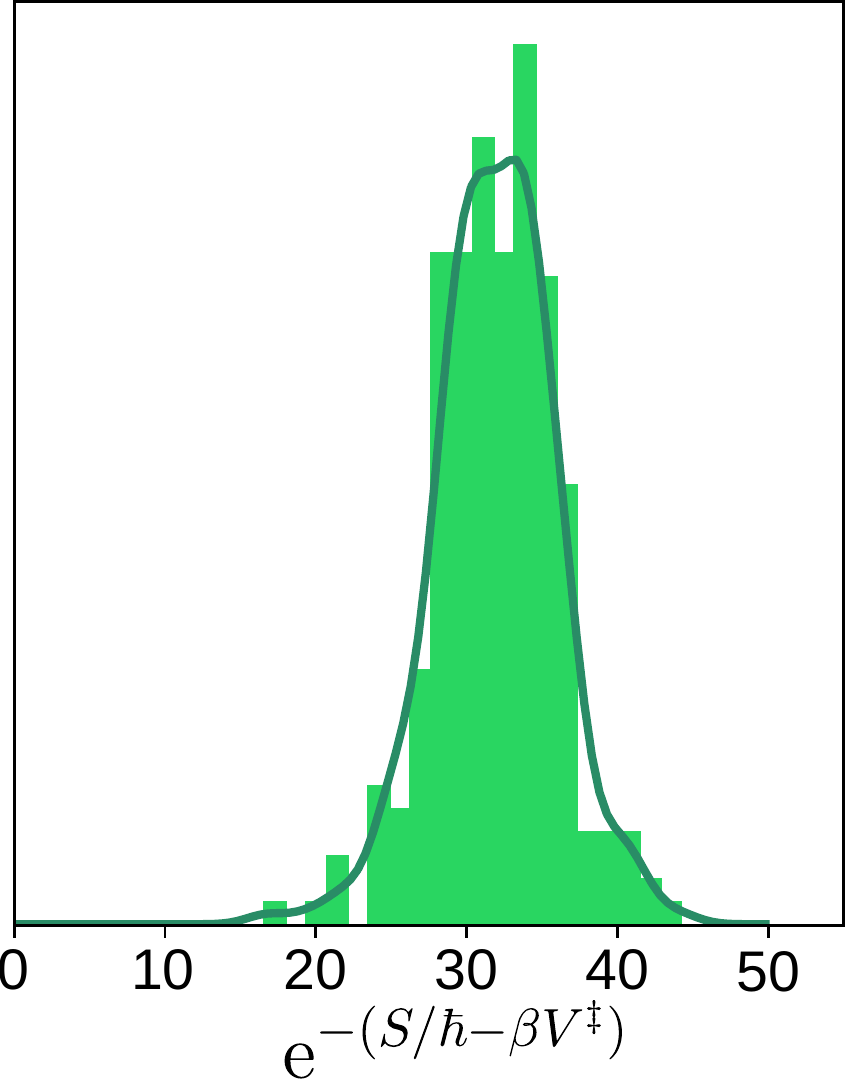}
    \caption{
    Distribution of the approximate tunnelling factor $\Gamma_{\text{inst}}$ of an ensemble of 250 golden-rule instantons.
    }
    \label{inst_dist}
\end{figure}
Fig.~\ref{inst_dist} shows the distribution of $\Gamma_{\text{inst}}$ of the instanton we computed. 
The average $\Gamma_{\text{inst}}$ of the ensemble is 32 (with a standard deviation of 4), suggesting that tunnelling plays a significant role in this system.

\section{Spin-boson model mapping}
\begin{figure}[H]
    \centering
    \includegraphics[width=0.66\columnwidth]{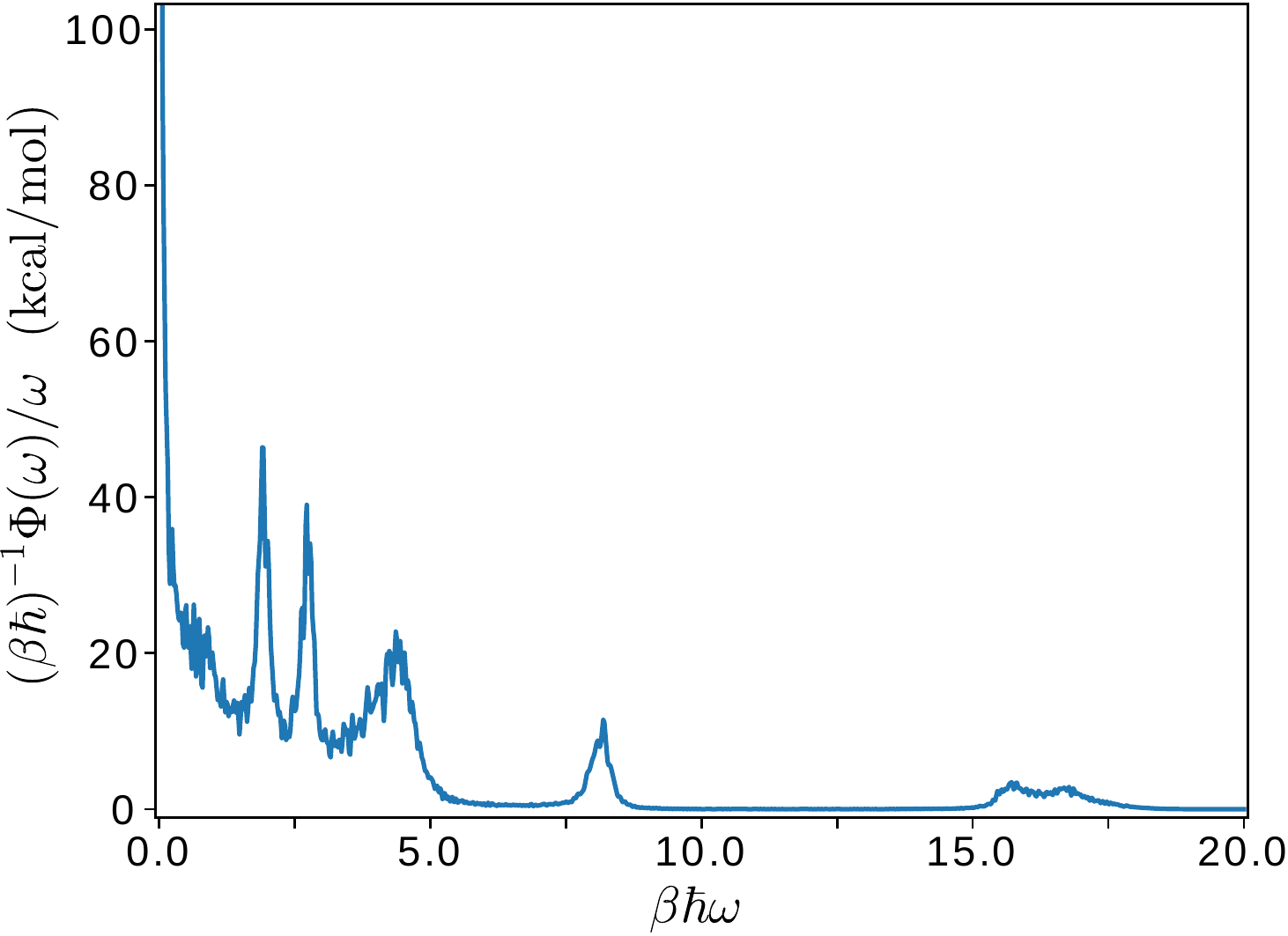}
    \caption{
    Spectral density of the ferrous--ferric system used to define
    the spin-boson model.
    $\Phi(\omega)/\omega$ is obtained from Fourier transform of the energy gap autocorrelation function calculated from classical NVE simulations.
    The frequency is measured in reduced units of $\beta\hbar\omega$,
    where $(\beta\hbar)^{-1}=208.5$ cm$^{-1}$ at the simulation temperature of 300 K.
    }
    \label{sb_spec}
\end{figure}

We mapped the ferrous--ferric system in q-TIP4P/F water onto a spin-boson model, using the same procedure described in Ref.~\cite{Bader}.
The spectral density is given in Fig.~\ref{sb_spec}.
Compared to the SPC water results in Ref.~\cite{Bader}, the positions of the three low-frequency peaks agree very well.
The highest of the three low-frequency peaks for q-TIP4P/F water is lower than SPC water (especially for the peak at $\beta\hbar\omega=4.3$), due to the lower reorganisation energy of the system with q-TIP4P/F water.
The high-frequency peaks originate from hydrogen bond bending and stretching, which are absent in the rigid SPC water. 


%